\RequirePackage[left, pagewise, displaymath, mathlines]{lineno}
\documentclass[aps,prx,twocolumn,superscriptaddress,groupedaddress,balancelastpage]{revtex4}
\usepackage[compat=1.0.0]{tikz-feynman}

\usepackage{amssymb}
\usepackage{multirow}
\usepackage{bm}
\usepackage{amsmath}
\usepackage{graphicx}
\usepackage{epstopdf}
\usepackage{subfigure}
\usepackage{natbib}
\usepackage{epsfig}
\usepackage{amsfonts}
\usepackage{mathrsfs}
\usepackage[toc,page,title,titletoc,header]{appendix}
\usepackage[colorlinks,urlcolor=black, linkcolor=blue,citecolor=blue,anchorcolor=blue]{hyperref}
\usepackage{dsfont,amsthm,amsbsy}

\def\tr{{\rm tr}}

\usepackage{fancyhdr}
\usepackage{braket}

\usetikzlibrary{quotes,angles}
\newcommand{\obtuseangle}{\kern.08em
\begin{tikzpicture}
    \draw coordinate (a) at (0.14,0);
    \draw coordinate (b) at (0,0);
    \draw coordinate (c) at (-.12,0.18);
    \draw (a) -- (b) -- (c) pic [draw=black]{} ;
\end{tikzpicture}
\kern.08em
}

\begin{document}

\title{Pair-Density-Wave in the Strong Coupling Limit of the Holstein-Hubbard model}
\author{Kevin S. Huang}
 \affiliation{Department of Physics, Stanford University, Stanford, California 94305, USA}

\author{Zhaoyu Han}
 \affiliation{Department of Physics, Stanford University, Stanford, California 94305, USA}

\author{Steven A. Kivelson}
\email{kivelson@stanford.edu}
  \affiliation{Department of Physics, Stanford University, Stanford, California 94305, USA}

\author{Hong Yao}
\email{yaohong@tsinghua.edu.cn}
 \affiliation{Institute for Advanced Study, Tsinghua University, Beijing 100084, China}
 \affiliation{Department of Physics, Stanford University, Stanford, California 94305, USA}

\date{\today}

\begin{abstract}
A pair-density-wave (PDW) is a novel superconducting state with an oscillating order parameter. A microscopic mechanism that can give rise to it has been long sought but has not yet been established by any controlled calculation. Here we report a density-matrix renormalization group (DMRG) study of an effective $t$-$J$-$V$ model, which is equivalent to the Holstein-Hubbard model in a strong-coupling limit, on long two-, four- and six-leg triangular cylinders. While a state with long-range PDW order is precluded in one dimension, we find strong quasi-long-range PDW order with a divergent PDW susceptibility as well as spontaneous breaking of time-reversal and inversion symmetries. Despite the strong interactions, the underlying Fermi surfaces and electron pockets around the $K$ and $K^\prime$ points in the BZ can be identified. We conclude that the state is valley-polarized and that the PDW arises from intra-pocket pairing with an incommensurate center of mass momentum. In the two-leg case, the exponential decay of spin correlations and the measured central charge $c\approx 1$ are consistent with an unusual realization of a Luther-Emery liquid.
\end{abstract}

\maketitle

In a PDW state, the Cooper pairs have non-zero center of mass momentum~\cite{agterberg2020physics}. Long-studied versions of such states are the Fulde-Ferrell-Larkin-Ovchinnikov (FFLO) phases~\cite{larkin1969quasiclassical,fulde1964superconductivity} that can arise in the context of BCS theory in the presence of a small degree of spin magnetization. However, it has been conjectured that a PDW could also arise from a strong-coupling mechanism and could be responsible for the dynamical layer decoupling phenomena observed in certain cuprate materials~\cite{himeda2002stripe,berg2007dynamical,berg2009striped}.  While PDW states have been shown to be energetically competitive in certain mean-field and variational calculations \cite{Moore2012prb, PhysRevX.4.031017, PhysRevB.89.165126, Jian2015prl}, and evidence has been sought in numerical studies of generalized $t$-$J$~\cite{himeda2002stripe, PhysRevB.76.140505, corboz2014competing, dodaro2017intertwined, PhysRevLett.122.167001} and Hubbard models~\cite{PhysRevLett.107.187001,Venderleyeaat4698}, PDW long-range order has not been established in any controlled calculation in two or higher dimensions. In one-dimensional systems - which can be reliably treated using DMRG - the closest one can get to such order is PDW quasi-long-range order with a divergent susceptibility.  However, even this has only been seen to date for the case of commensurability-two PDW order in Heisenberg-Kondo chains~\cite{PhysRevLett.105.146403, PhysRevB.85.035104}; other reported DMRG sightings~\cite{peng2020evidence, PhysRevLett.122.167001,Hongchentriangle} have found PDW correlations that fall off fast enough to yield only finite superconducting susceptibility.

\begin{figure}[b!]
	\centering
	\subfigure[]{\label{fig:lattice}\includegraphics[width=0.40\linewidth]{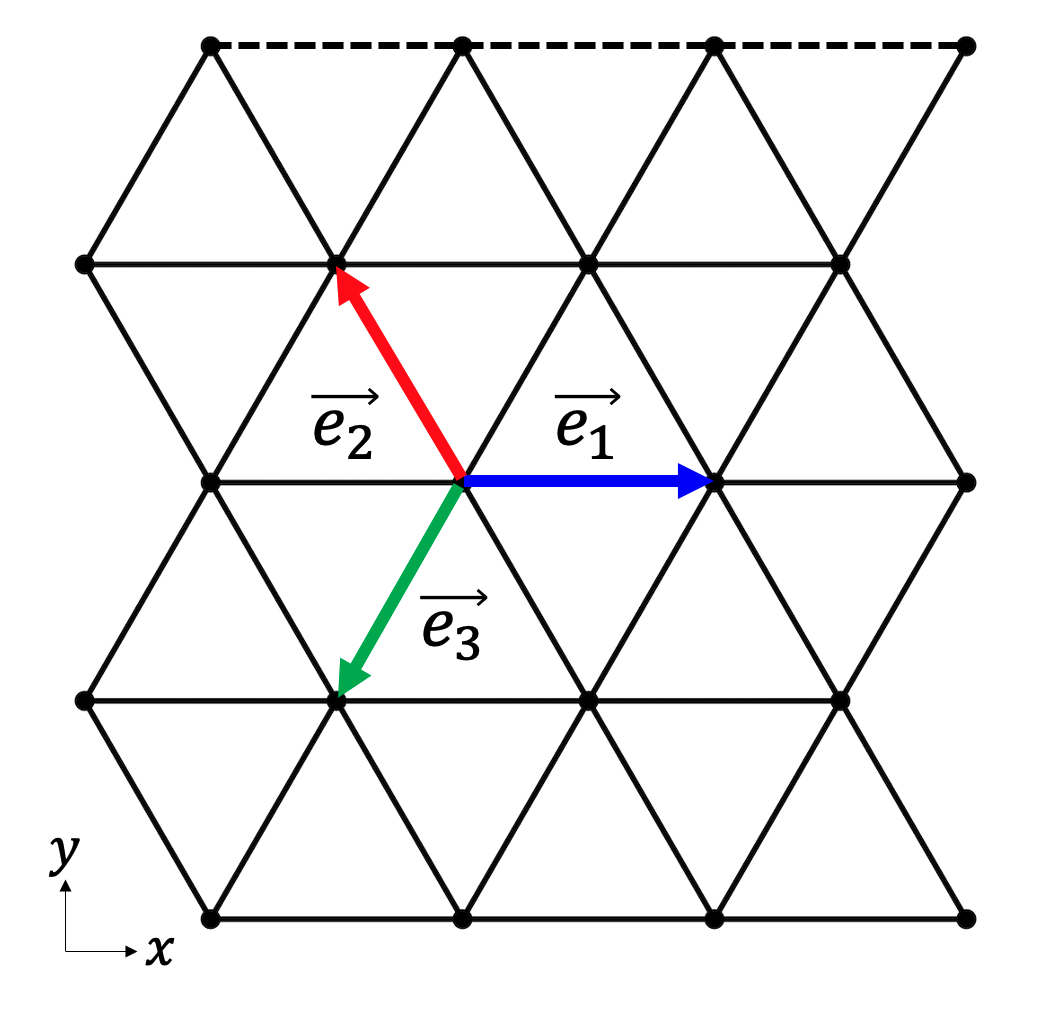}}
	\subfigure[]{\label{fig:BZ}\includegraphics[width=0.48\linewidth]{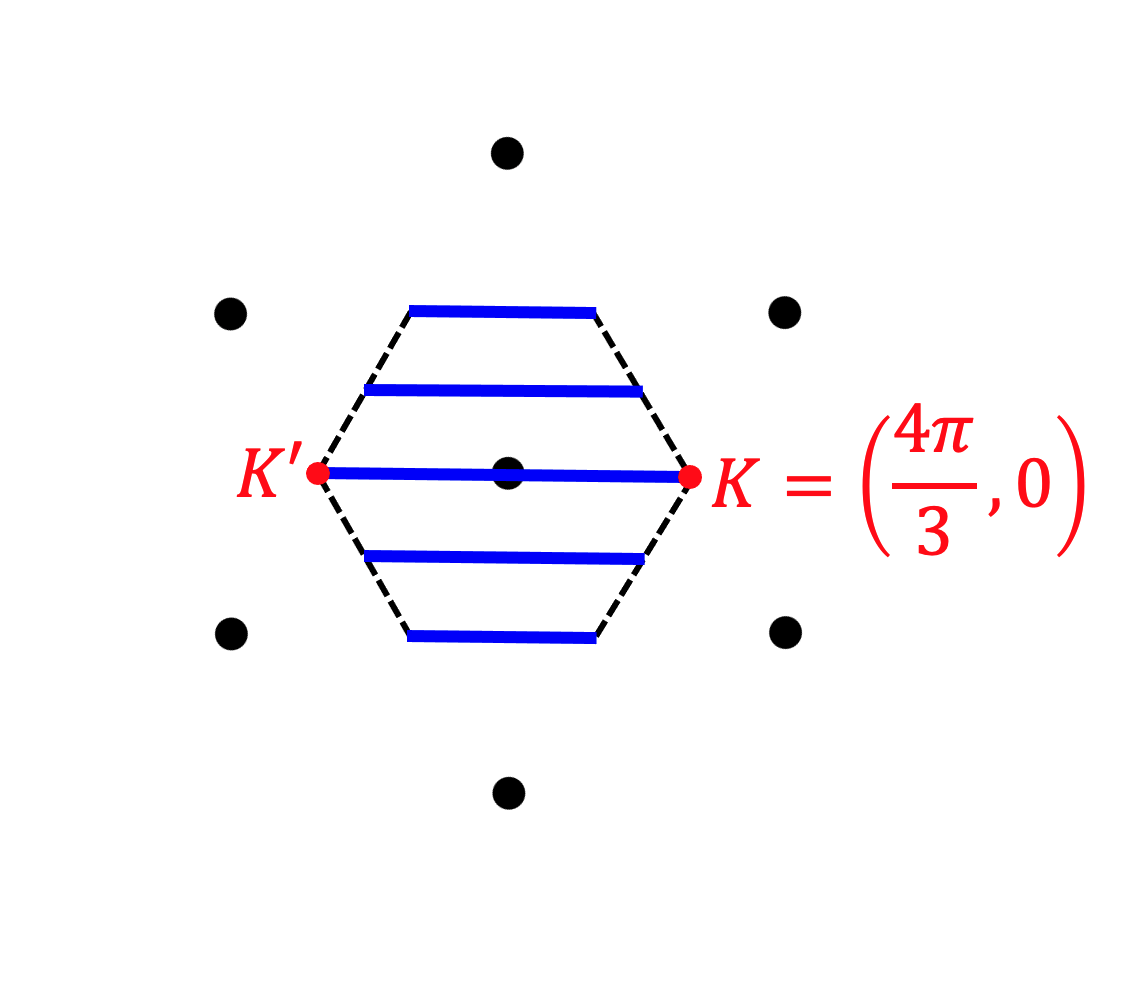}}
	\caption{(a) An illustration of the four-leg triangular cylinder (identifying top row of sites with bottom row of sites). Three primitive vectors are labeled on the nearest-neighboring bonds of a site. (b) The corresponding first BZ is contained within the black dashed lines with $K$, $K'$ points marked. The blue bands indicate allowed values of momenta for the four-leg cylinder.}
\end{figure}

Previously~\cite{PhysRevLett.125.167001}, we analyzed a strong coupling limit of the Holstein-Hubbard model, where the on-site direct electron-electron repulsion, $U_\text{e-e}$, is stronger than the phonon induced attraction, $U_\text{e-ph}$. We showed that the low energy physics is captured by an effective model of a form similar to the $t$-$J$-$V$ model derived from the strong coupling limit of the ordinary Hubbard model, but in an unfamiliar range of parameters. In the effective model, the electron hopping is exponentially suppressed by a Frank-Condon factor, $\mathrm{e}^{-\frac{U_{\text{{e-ph}}}}{2\omega_D}}$, relative to its bare value, $t$, where $\omega_D$ is the optical phonon frequency. Meanwhile, the nearest-neighbor magnetic exchange coupling $J$ and density repulsion $V$, are relatively unsuppressed. Thus one can readily access novel regimes in which the effective interactions are comparable to or larger than the electron kinetic energy. Specifically, we analyzed the model on the two-dimensional triangular lattice in the adiabatic limit, where the suppression is large. We found a PDW when the electrons are dilute and the bare hopping integral $t$ is negative (i.e. the band {\em maximum} occurs at $\vec k=\vec 0$ while {\em minima} occur at the $K$ and $K'$ points). Unfortunately, in the adiabatic limit, the PDW is a Bose-Einstein condensate (BEC) of real space pairs, in which the same Frank-Condon factor suppresses the condensation temperature, meaning that this order would be destroyed at exponentially small temperatures.

\begin{figure}
	\subfigure[]{\label{fig:inphase}\includegraphics[width=0.93\linewidth]{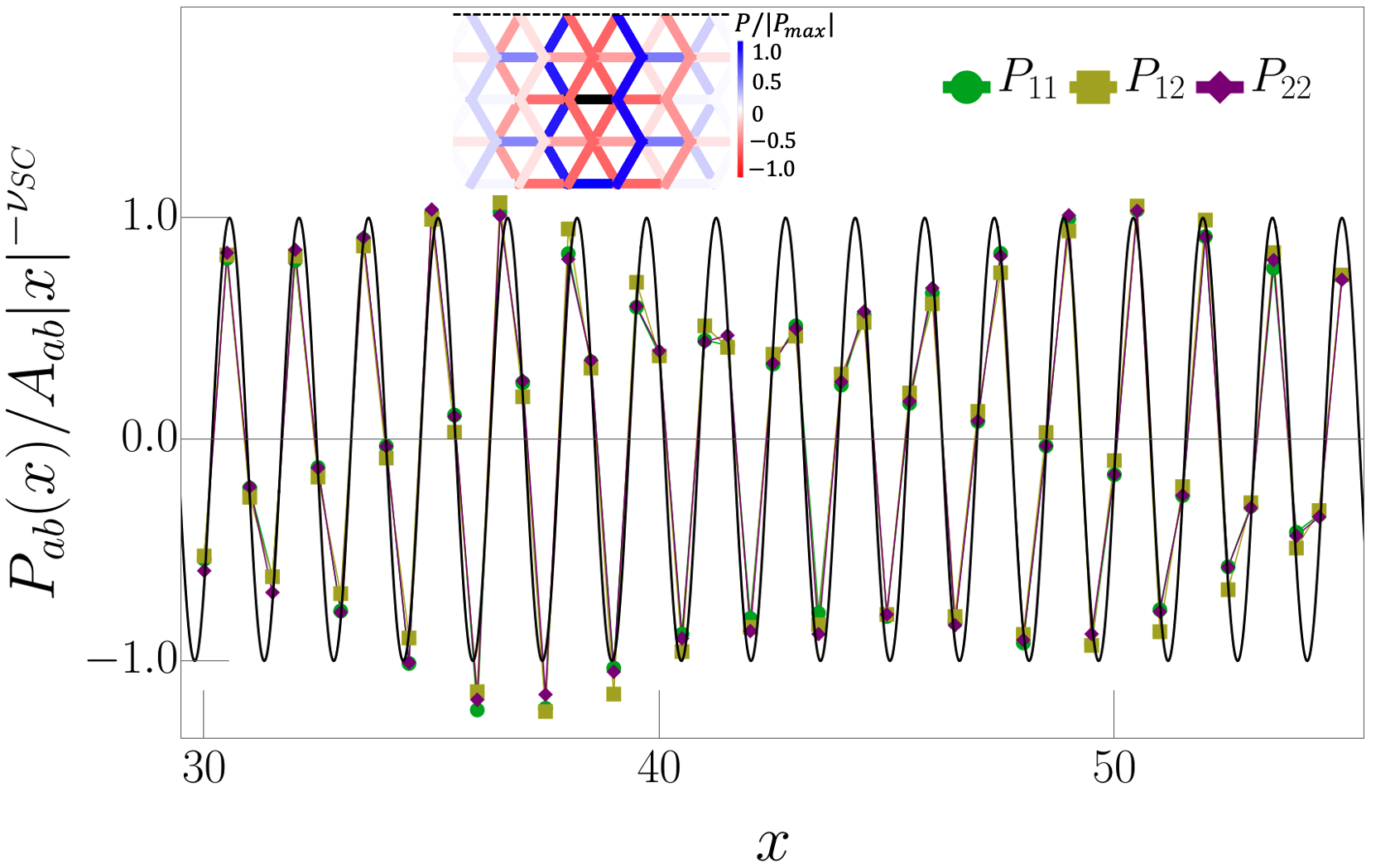}}
	\subfigure[]{\label{fig:Spdw_peaks}\includegraphics[width=0.93\linewidth]{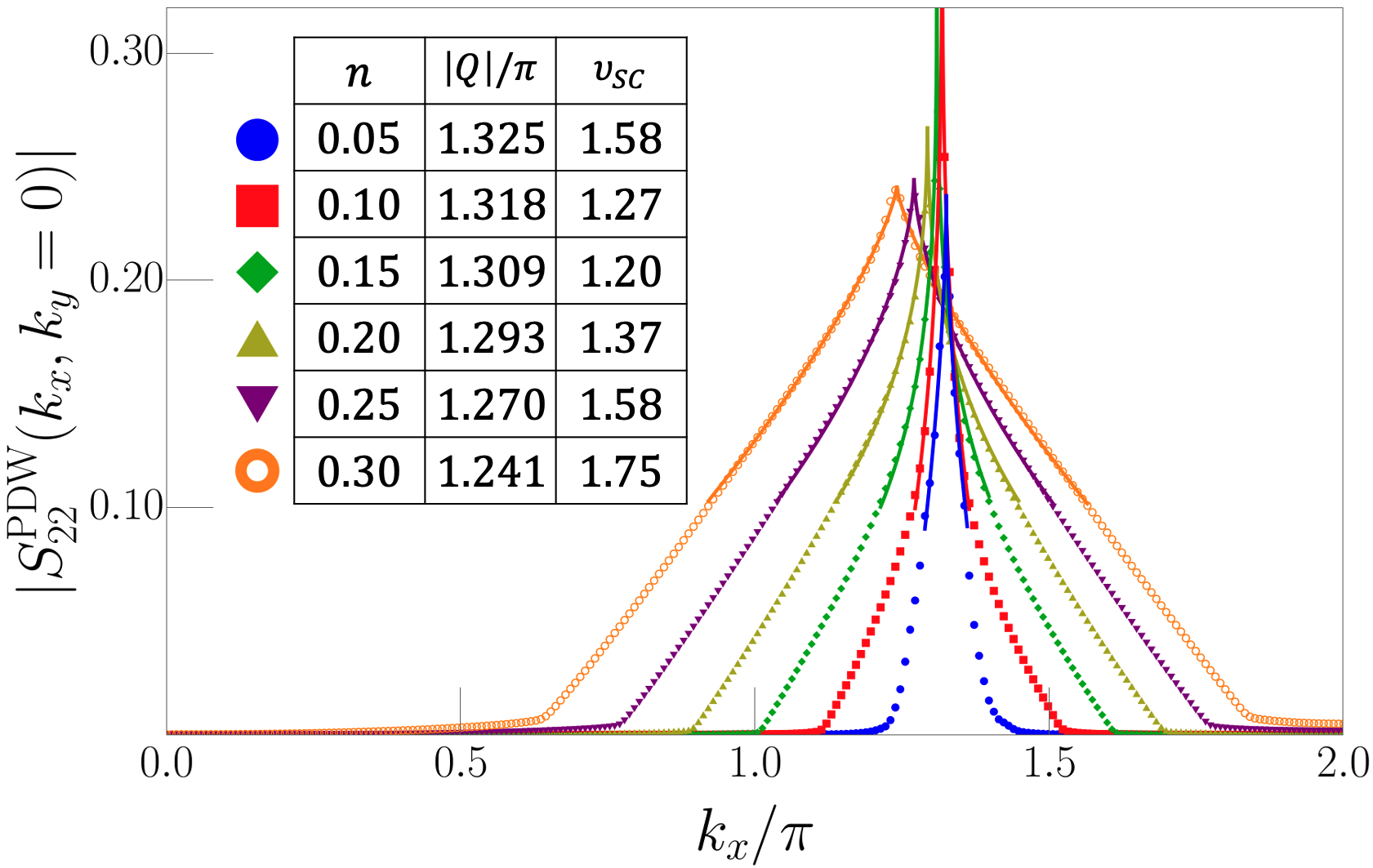}}
	\subfigure[]{\label{fig:Spdw_legs}\includegraphics[width=0.93\linewidth]{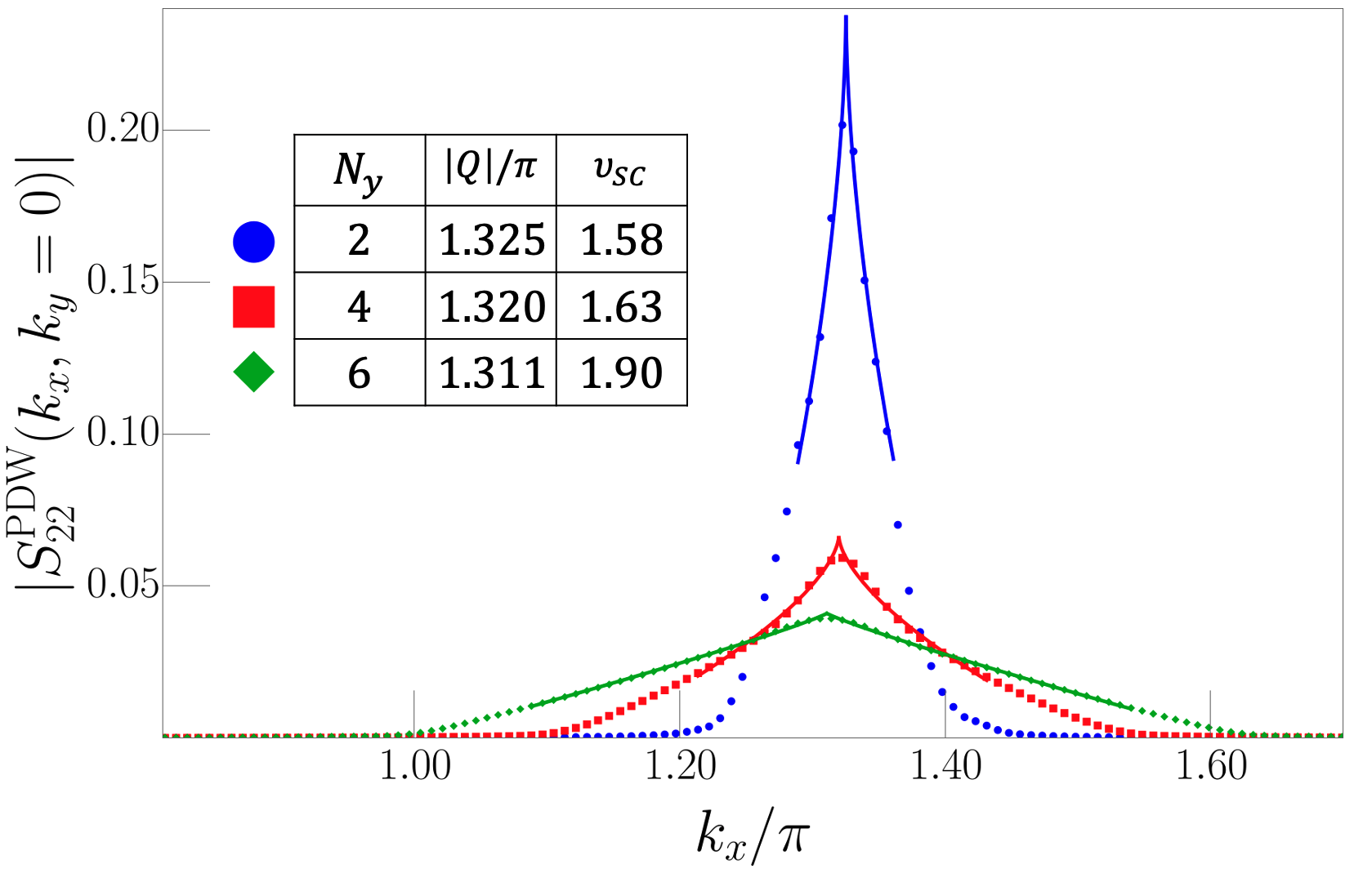}}
	\caption{\label{fig:2} (a) The pair-pair correlator normalized by a power-law decay function $A_{ab}x^{-\nu_\text{SC}}$ with $\nu_{SC}=1.20$ for a two-leg cylinder of length $N_x=240$ and electron density $n=0.15$. The black reference curve is $\cos(Qx+\phi_\text{SC})$ with $Q=1.309\pi$ extracted from the structure factor. The inset shows the short-distance pair correlation pattern around the center of a four-leg cylinder of length $N_x=100$ and electron density $n=0.15$ with the reference bond colored in black. (b) Magnitude of structure factor $S^\text{PDW}_{22}(k_x,k_y=0)$ at various electron densities on the two-leg cylinder of length $N_x=240$. The interval $k_x \in (-2\pi,0)$ mirrors the range shown. The fitted curves with functional form $A - B |k_x-Q|^{{\nu_\text{SC}}-1}$ are plotted as solid lines with the same color for each filling, and the fitted peak positions $Q$ and exponents $\nu_{\text{SC}}$ are listed. (c) Magnitude of structure factor for two-, four- and six-leg cylinders with $n=0.05$ and correspondingly $N_x=240,100,60$.}
\end{figure}

\begin{figure}[t!]
    \centering
    \subfigure[]{\label{fig:currentcorrelation}\includegraphics[width=0.99\linewidth]{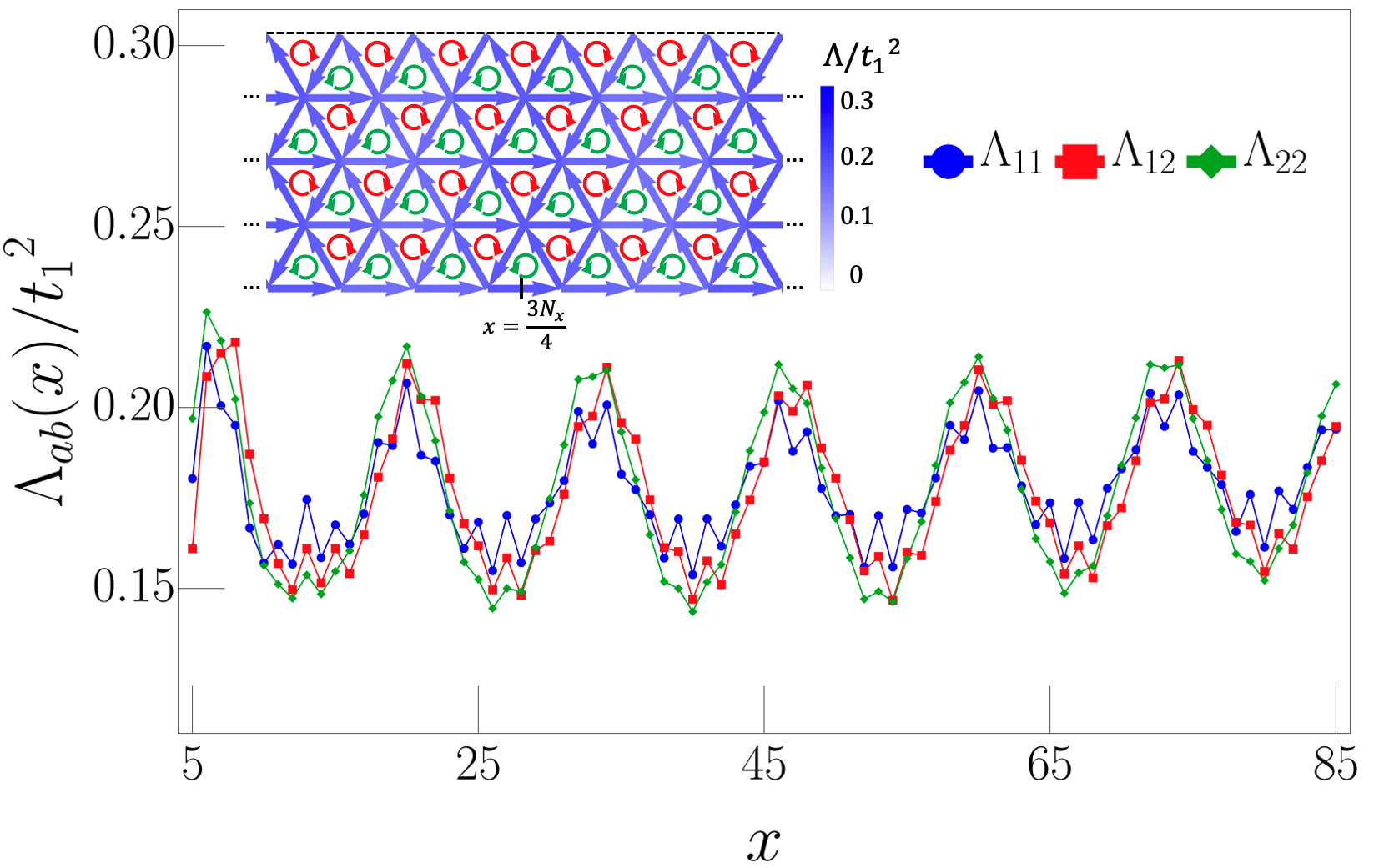}}
    \subfigure[]{\label{fig:fermidensity}\includegraphics[width= 0.46\linewidth]{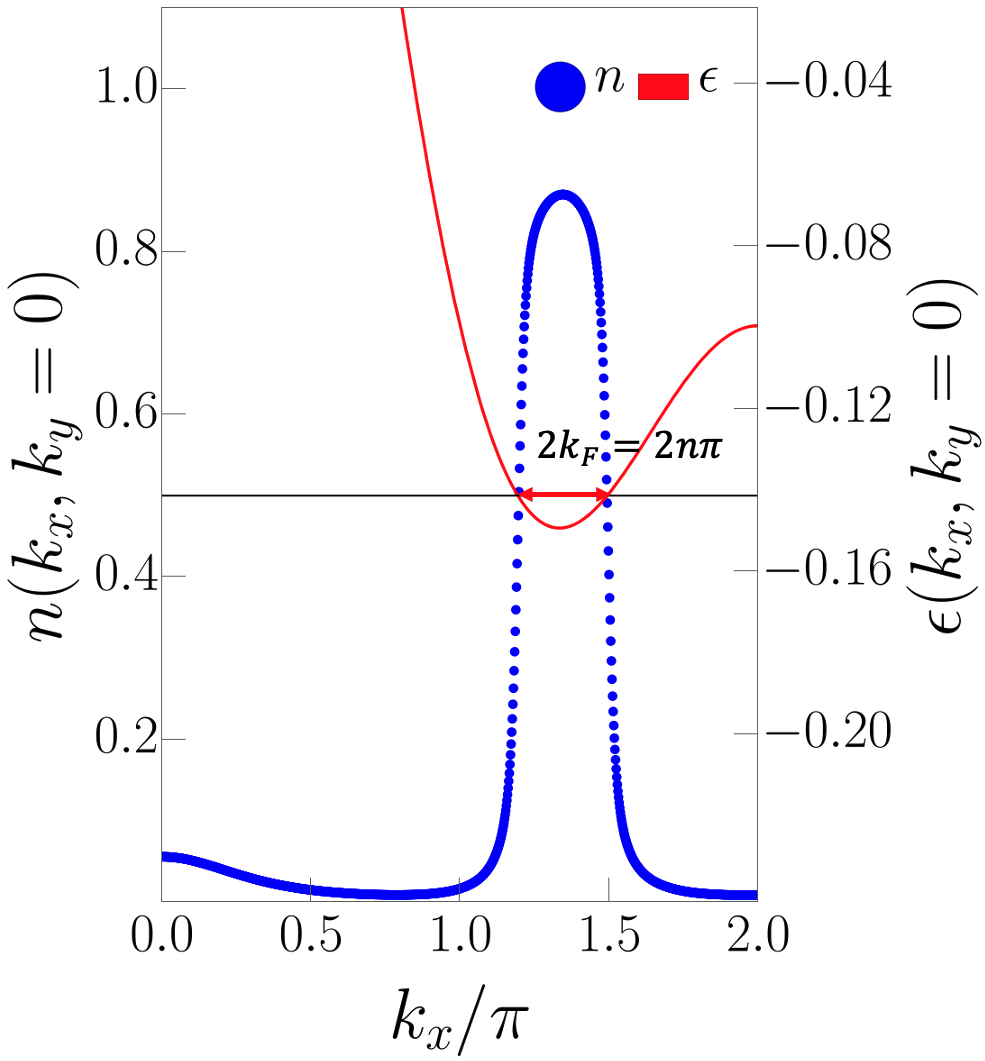}}
    \subfigure[]{\label{fig:Qfit}\includegraphics[width=0.44\linewidth]{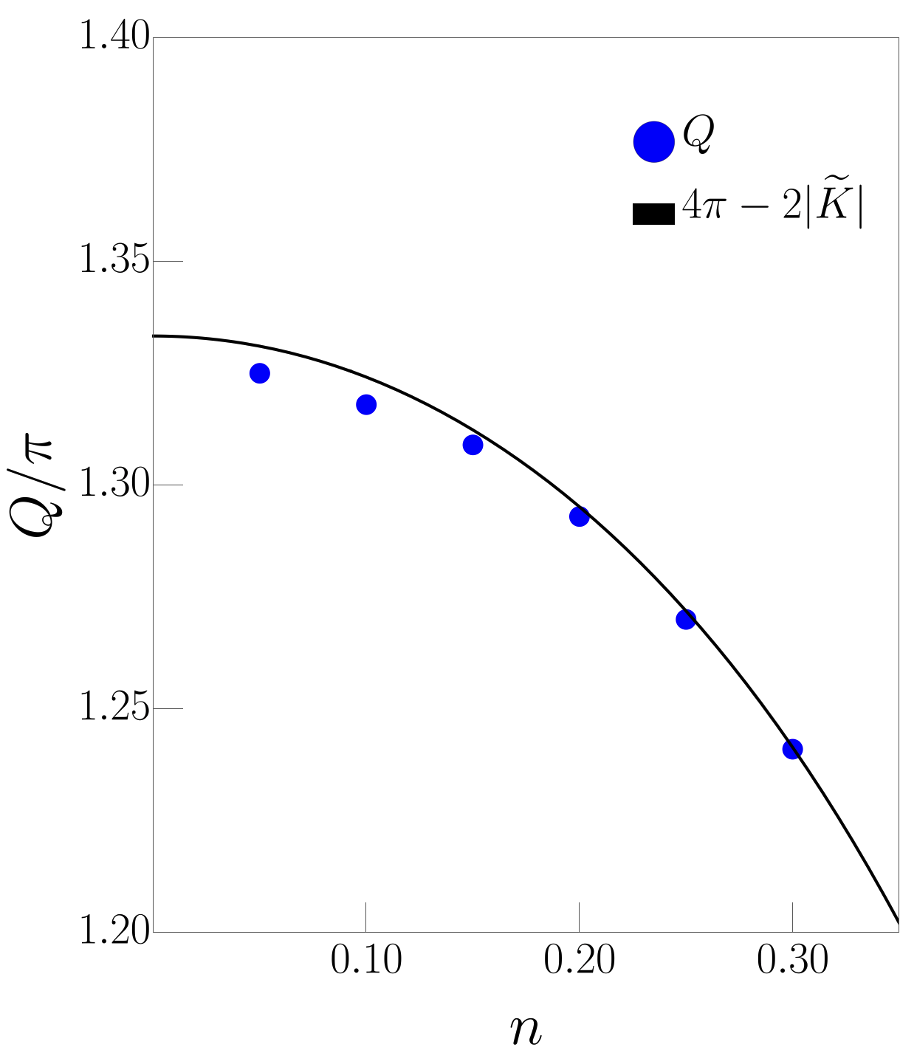}}

    \caption{  (a) The current-current correlator for a two-leg cylinder of length $N_x=240$ and electron density $n=0.15$. The inset shows the long-distance current correlation pattern around $x=3N_x/4$ of a four-leg cylinder of length $N_x=100$ and electron density $n=0.15$, with reference bond at $x=N_x/4$. (b) The occupation number in momentum space of a two-leg cylinder with length $N_x=240$ and electron density $n=0.15$. The Fermi points match those of the non-interacting band $\epsilon(\vec{k})$  defined in Eq.~\ref{dispersion} with width $2k_F=2n\pi=0.3\pi$. The interval $k_x \in (-2\pi,0)$ mirrors the range shown. (c) Despite strong interactions, there is good agreement between the extracted pair momentum listed in Fig.~\ref{fig:Spdw_peaks} and the center of the non-interacting band structure estimated in Eq.~(\ref{Ktilde}). }
\end{figure}

In the present work,  we employ DMRG to study the effective model in a range of parameters suggested by the above discussed analysis, but now for the case of a finite Frank-Condon factor \footnote{The range of $V/J$ over which PDW order occurs was found to be $[0.43,1]$  in the adiabatic limit in \cite{PhysRevLett.125.167001}.  From DMRG we find that that this range shifts somewhat with increasing Frank-Condon overlap factor.  Nonetheless, the range in the adiabatic limit provides guidance in  searching PDW, and indeed the data shown in this paper  lies this range. }.  For all the cylinders we can treat in this way, we find a divergent PDW susceptibility, as shown in Fig.~\ref{fig:2}. The PDW state spontaneously breaks time-reversal and inversion symmetries, resulting in a local pattern of equilibrium currents as shown in the inset of Fig.~\ref{fig:currentcorrelation}. The state is far from the BEC limit as can be seen in Fig.~\ref{fig:fermidensity} from the existence of a sharp drop in the electron occupation at well-defined Fermi points corresponding  to electron pockets around the $K$ and $K'$ points in the  Brillouin Zone (BZ). However, the size of the pockets (i.e. the value of $2k_F$) is a factor of two larger than what it would be for non-interacting electrons, which we will argue reflects spontaneous valley polarization. As shown in Figs.~\ref{fig:inphase}\&\ref{fig:Qfit}, the PDW ordering wavevector, $Q$, is incommensurate and density dependent. We will show that the ordering vector takes the value expected for intra-pocket pairing.

{\bf Model and Methods.} - The effective Hamiltonian (with implicit projection to states with no doubly occupied sites) is:
\begin{linenomath}
\begin{align}\label{t-J-V}
\hat{H}_{\text{eff}} =& - \Bigg \{t_1 \sum_{\langle i,j\rangle,\sigma} \hat{c
}_{i,\sigma}^{\dagger}  \hat{c}_{j,\sigma}   +  t_2 \sum_{\langle i, m, j\rangle,\sigma}  \hat{c}_{i,\sigma}^{\dagger} (1-2\hat n_m) \hat{c}_{j,\sigma} \nonumber \\
& + (\tau+2t_2)\sum_{\langle i, m, j\rangle} \hat{s}^{\dagger}_{im} \hat{s}_{mj} + {\rm h.c.} \Bigg\}
\nonumber \\
& \  + J\sum_{\langle i,j\rangle }\left[\Vec{{S}}_i\cdot\Vec{{S}}_j - \frac{\hat{n}_i \hat{n}_j}{4} \right] + V \sum_{\langle i,j\rangle} \hat{n}_{i} \hat{n}_{j}
\end{align}
\end{linenomath}
where $t_1$ is the renormalized nearest-neighbor hopping, $t_2$ is a weak next-nearest-neighbor hopping term via an intermediate site $m$, $\langle i,m,j\rangle$ represents a triplet of sites such that $m$ is a nearest-neighbor of two distinct sites $i$ and $j$, and $(\tau+2t_2)$ is a weak singlet hopping term where $\hat{s}_{ij} = (\hat{c}_{i,\uparrow} \hat{c}_{j,\downarrow}+\hat{c}_{j,\uparrow} \hat{c}_{i,\downarrow})/\sqrt{2}$ is the annihilation operator of a singlet Cooper pair on bond $\langle ij\rangle$. Explicit expressions are given in Ref. \cite{PhysRevLett.125.167001} for the values of these effective couplings as functions of the parameters in the original Holstein-Hubbard model. For the data shown in this study, we fix $t=-1$, $\omega_D=3$, $U_\text{e-e}=22$, and $U_\text{e-ph}=18$, which leads to $J \approx 0.208$, $ V/J \approx 0.666$, $t_1/J \approx -0.240$, $t_2/J \approx 0.0329$ and $\tau/J \approx 0.0445$ in the effective model. However, we have checked that in a broad range of parameters, or setting $t_2=\tau =0$, the results we find do not change qualitatively. We include some results with different settings in the Supplemental Material.

We perform DMRG on triangular cylinders with $N_x \times N_y$ sites with open boundary conditions in the $x$-direction and periodic boundary conditions in the $y$-direction. The $x$-direction is aligned with a primitive vector, as shown in Fig.~\ref{fig:lattice}.  $\hat{e}_{a=1}=(1,0)$, $\hat{e}_{a=2,3}=(\frac{1}{2},\pm \frac{\sqrt{3}}{2})$ are the three primitive vectors of the triangular lattice. The compactification of the lattice to a cylinder restricts us to $N_y$ even. The momentum values in the $y$-direction take values $k_y = \frac{m}{N_y}\frac{4\pi}{\sqrt{3}}$, $m\in \mathbb{Z}$, indicated in Fig.~\ref{fig:BZ}. We will primarily focus on the two-leg case but some key findings are also verified in the four- and six-leg cases. Note that two-leg cylinders can be flattened to be a purely one-dimensional chain by simply neglecting the $y$ coordinate of each site, so we plot the data for all sites in the two-leg case. All the DMRG data collected are obtained from the lowest energy state out of five trials with independently randomized initial states and all the results shown (unless otherwise stated) are extrapolated to zero truncation error, utilizing data collected with six truncation errors ranging from $3\times 10^{-6}$ to $1\times 10^{-7}$. In the two-leg case, we have checked our results do not change significantly down to truncation error $7 \times 10^{-11}$, corresponding to keeping bond dimensions up to $m=4320$. The correlation functions shown are averaged over all legs of the ladder, taking $N_0=5N_y$ different reference sites $\vec{r}_0$ centered around the middle of the system. All data involving sites within $N_x/4$ to the open boundary are discarded, i.e. we only retain the data on the interval $x\in [N_x/4, 3N_x/4]$, to reduce boundary effects. Thus, each structure factor we compute utilizes the data on $N=N_xN_y/2$ sites.

{\bf Pair density wave correlations.} - We first examine the equal-time pair-pair correlations. The singlet pairing order parameter is defined on the nearest-neighboring bonds to a given site $\vec{r}_i$ as $\Delta_a(\vec{r}_i) \equiv \hat{c}_{\vec{r}_i, \uparrow}\hat{c}_{\vec{r}_i+\hat{e}_a, \downarrow}+\hat{c}_{\vec{r}_i+\hat{e}_a, \uparrow}\hat{c}_{\vec{r}_i, \downarrow}$. The pair-pair correlator is defined as $P_{ab}(\Vec{r}) \equiv \frac{1}{N_0} \sum_{\Vec{r}_0} \langle \Delta^{\dagger}_a(\Vec{r}+\Vec{r}_0)\Delta_b(\Vec{r}_0)\rangle$. Our principle finding is that $P_{ab}(\vec r)$ oscillates in sign and falls slowly in magnitude at large distance, i.e.
\begin{linenomath}
\begin{align}
    P_{ab}(\Vec{r}) \sim A_{ab}|x|^{-\nu_\text{SC}} \cos( Q x+\phi) .
    \label{pdwofr}
\end{align}\end{linenomath}
This can be seen in Fig.~\ref{fig:inphase}, which is a plot of $P_{ab}(\vec r)/A_{ab}|x|^{-\nu_{\text{SC}}} $ at large distance for a two-leg cylinder, taking ${\nu_\text{SC}}=1.20$. Without breaking of the reflection symmetry across the $x$ direction, $\hat{e}_{2}$ and $\hat{e}_3$ are equivalent, so we only show pairing correlations involving $\Delta_1$ and $\Delta_2$.  The fact that all the components of $P_{ab}$  oscillate in-phase with the same wavevector, $Q$,  indicates the local s-wave character of the PDW. An illustration of the pair correlation in the four-leg case can be seen in the inset of Fig.~\ref{fig:inphase}, which matches our previous prediction in Ref.~\cite{PhysRevLett.125.167001} at short distance.

Correspondingly, the Fourier transform of the correlator, i.e. the structure factor $S^{\text{PDW}}_{ab}(\vec{k})\equiv \frac{1}{N}\sum_{\Vec{r}_i,\vec{r}_j}\mathrm{e}^{\mathrm{i}\vec{k}\cdot (\vec{r}_i-\vec{r}_j)}\langle \Delta^{\dagger}_a(\Vec{r}_i)\Delta_b(\Vec{r}_j)\rangle$, has pronounced peaks at two non-zero momenta and vanishes at zero momentum, as shown in Fig.~\ref{fig:Spdw_peaks}. Around each peak but outside a cutoff window of width $\delta k_x \sim \frac{2\pi}{N_x}$ (to avoid finite size effects), the structure factor is well-fitted by a functional  form $A - B |k_x-Q|^{{\nu_\text{SC}}-1}$ as is expected, given the long-distance correlation behavior in Eq. (\ref{pdwofr}). It is important to note that the wavevector of the PDW, $|Q|$, appears to be a smooth monotonic  function of the electron density and is not locked to a multiple of the momenta at the $K$ points, i.e. the PDW is incommensurate with the lattice. 

Since 1D systems generically exhibit emergent Lorenz invariance, we can infer from the equal time correlator that the static susceptibility to PDW order should vary as $\chi(Q) \sim \max[\frac{1}{N_x},T]^{-(2-\nu_\text{SC})}$ as the system size tends to infinity and temperature to zero~\cite{giamarchi2003quantum}. Thus, $\chi(Q)$ diverges at zero temperature for $\nu_\text{SC}<2$, as we find is the case for a wide range of electron densities and model parameters. That the divergence of susceptibility is not restricted to the two-leg case can be seen in Fig.~\ref{fig:Spdw_legs}, which shows the structure factor and the extracted exponents at the same electron density $n=0.05$ for the four and six-leg cases.

{\bf Spontaneously broken time-reversal and inversion symmetries.} - Another prominent feature of the ground state is that it spontaneously breaks both time-reversal and inversion symmetries. This can be directly seen from the current-current correlation $\Lambda_{ab}(\Vec{r}) \equiv \frac{1}{N_0} \sum_{\Vec{r}_0} \langle \Vec{J}_a(\Vec{r}+\Vec{r}_0)\cdot\Vec{J}_b(\Vec{r}_0)\rangle$, where $\Vec{J}_a(\Vec{r}_i) \approx - \mathrm{i} t_1 \hat e_a\ \sum_\sigma (\hat{c}^\dagger_{\Vec{r}_i,\sigma}\hat{c}_{\Vec{r}_i+\hat{e}_a,\sigma} -\hat{c}^\dagger_{\Vec{r}_i+\hat{e}_a,\sigma}\hat{c}_{\Vec{r}_i,\sigma})$ is the current on the bond directed in the $\hat{e}_a$ direction. (The actual current operator that we compute also receives minor contributions from $t_2$ and $\tau$, see Supplemental Materials for the full expression). As shown in Fig.~\ref{fig:currentcorrelation}, the current-current correlation oscillates around a non-zero value at long distance, and we also confirm in Supplemental Materials that the peak of the structure factor of the current-current correlation scales linearly with system size. These facts signify persisting currents in the ground states. The pattern of the current flows is shown in the inset, indicating that the ground state is an orbital anti-ferromagnet.

\begin{figure}[h!]
    \subfigure[]{\label{fig:chargecorrelation}\includegraphics[width=0.56\linewidth]{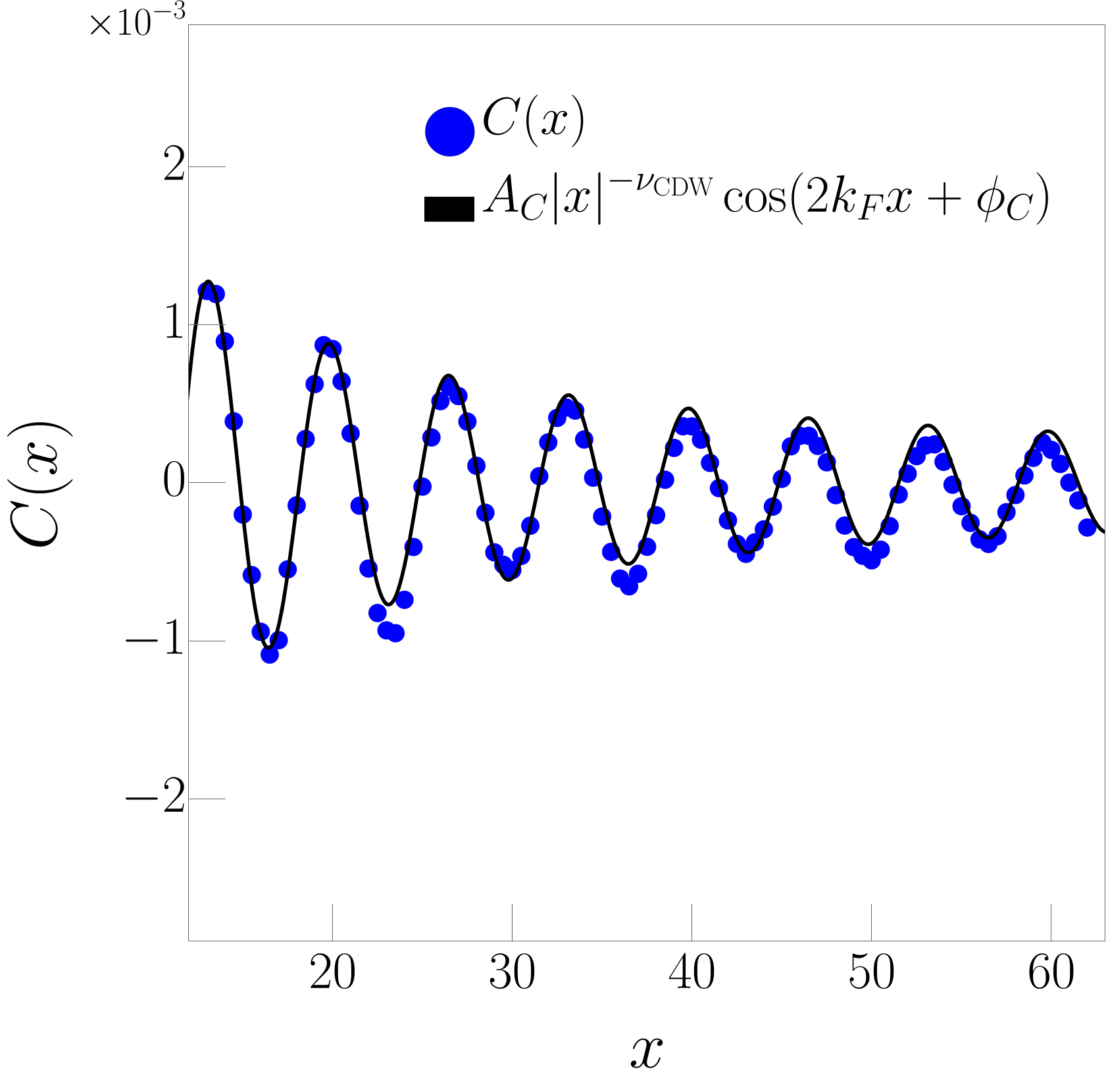}}
    \subfigure[]{\label{fig:spincorrelation}\includegraphics[width=0.4\linewidth]{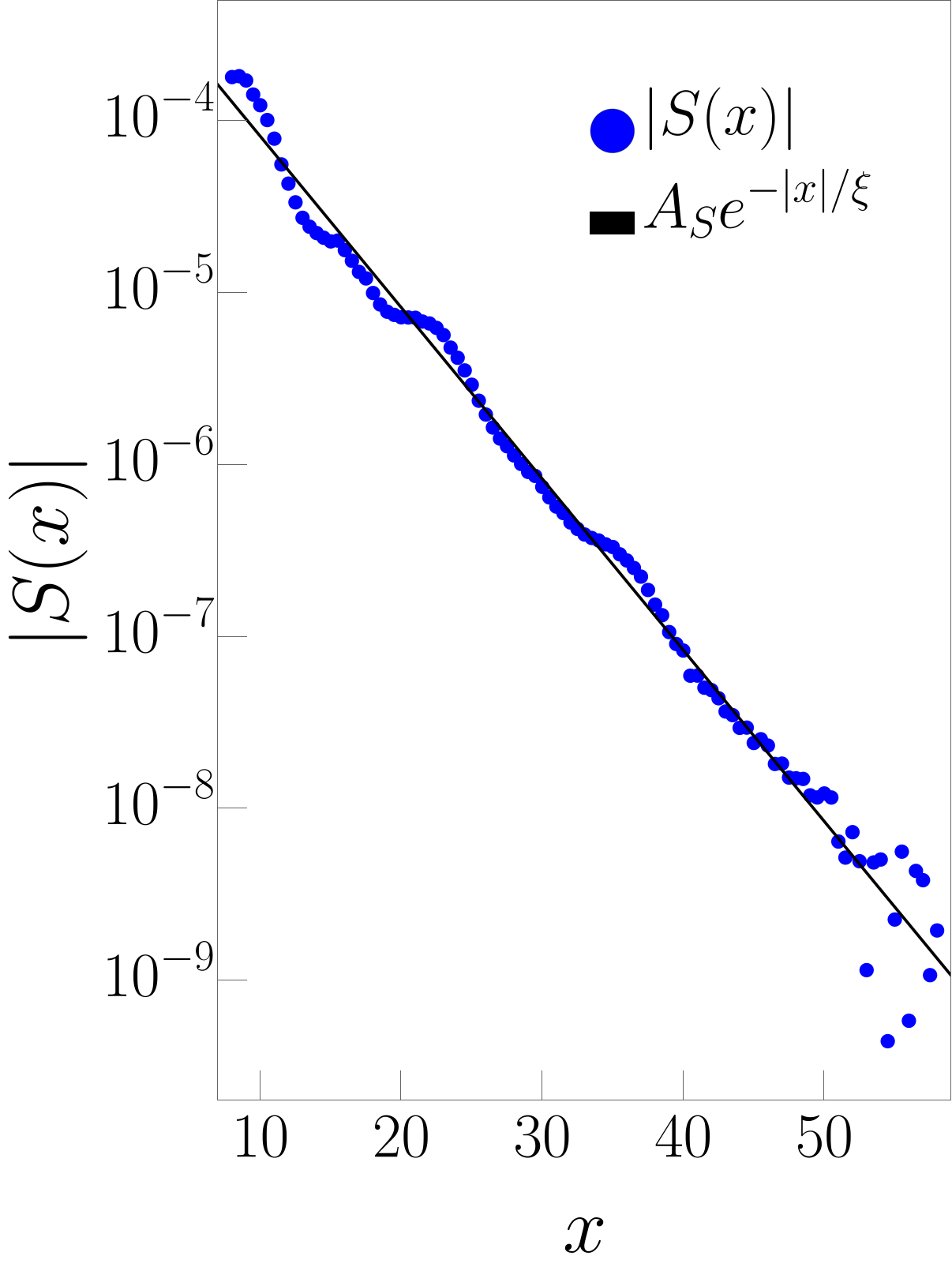}}
    \subfigure[]{\label{fig:centralcharge}\includegraphics[width=0.99\linewidth]{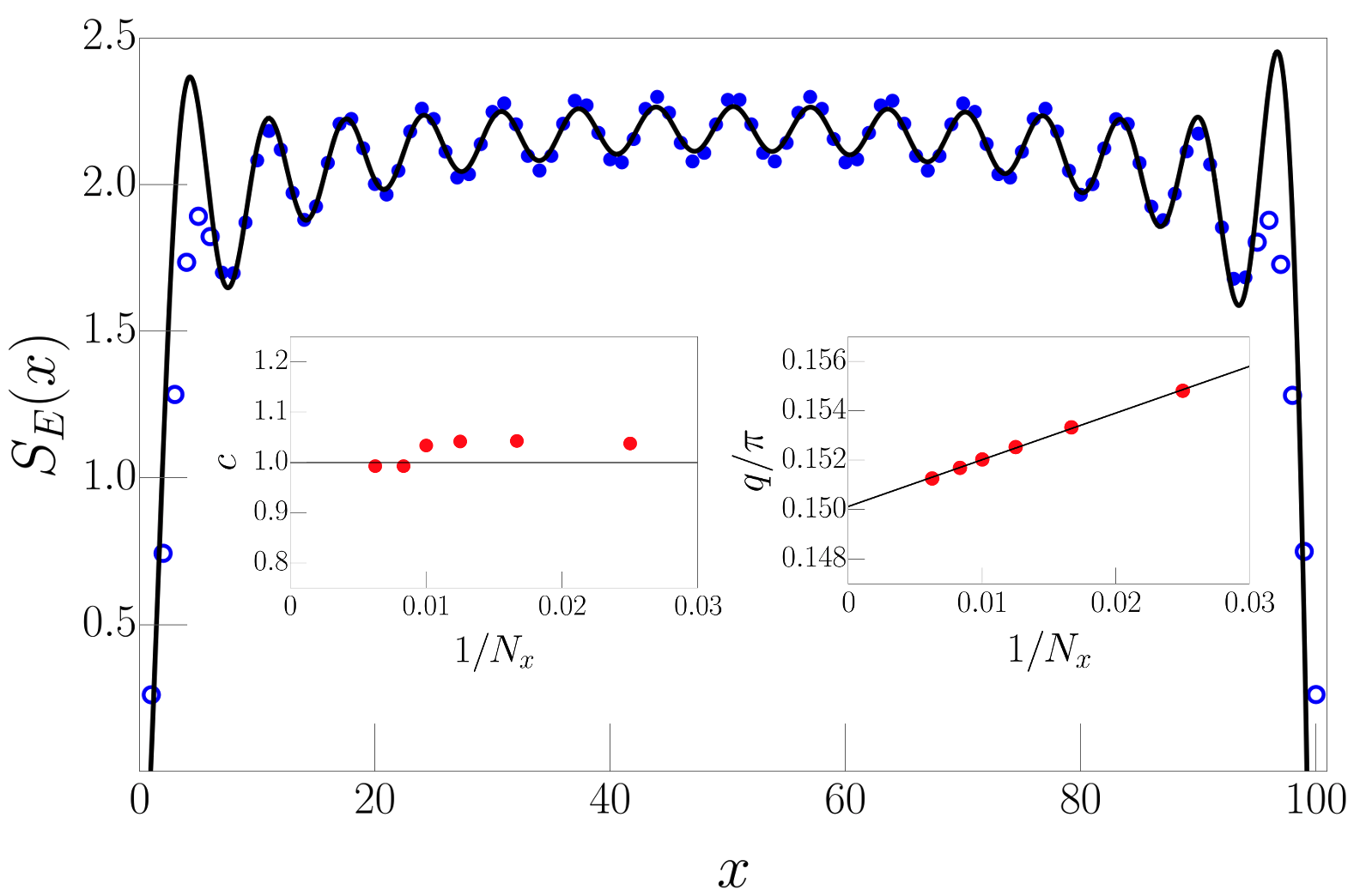}}
    \caption{(a) Real-space fitting $A_C|x|^{-\nu_{\text{CDW}}}\cos(2k_Fx+\phi_C)$ of the CDW correlation for a two-leg cylinder with $N_x=240$ and $n=0.15$ yields $\nu_{\text{CDW}}=0.90$. (b) Real-space fitting $A_Se^{-|x|/\xi}$ of SDW correlation for a two-leg cylinder with $N_x=240$ and $n=0.15$ yields $\xi=4.36$. (c) Von Neumann entanglement entropy for a two-leg cylinder with $N_x=100$ and $n=0.15$ (truncation error $5 \times 10^{-11}$). Discarding data points at the edges (indicated by open markers), the entropy is fit to Eq.\,\ref{entropyfit}. The extracted central charge is $c=1.03$. Left inset: extracted central charge across a range of lengths with $n=0.15$ agrees well with $c=1$. Right inset: linear extrapolation of the fitted $q$ in the thermodynamic limit yields $q=0.150\pi$ matching $k_F=n\pi=0.15\pi$.}
\end{figure}

{\bf Valley polarization.} - We further identify the broken symmetry states by investigating the occupation number in momentum space, $n(\Vec{k})\equiv \frac{1}{N}\sum_{\vec{r}_i,\vec{r}_j,\sigma}e^{i\vec{k} \cdot (\vec{r}_i-\vec{r}_j)}\braket{c_{\vec{r}_i,\sigma}^{\dagger}c_{\vec{r}_j,\sigma}}$, for the two-leg cylinder. As shown in Fig.~\ref{fig:fermidensity}, although the strong interactions shift a small fraction of occupation weight to the vicinity of zero momentum (far above the non-interacting Fermi surface), most of the weight is confined to narrow intervals of $k$ about the two minima of the non-interacting bands,  which occur at $K=(4\pi/3,0)$ and $K' =-K$. However, the width of these intervals - labeled ``$2k_F$'' in the figure - is twice as large as it would be for a non-interacting system. Rather, this observation is consistent with the supposition that the ground-state is valley polarized~\footnote{We have ruled out the possibility of a spin-polarized ferromagnet by checking that the ground-state has spin $0$ (shown in Fig.~\ref{fig:spincorrelation}), and the possibility of spin-valley locked polarization by confirming that the spin current correlations exponentially decay in space (shown in Supplemental Materials). We also checked that the triplet pairing correlations are weak and extremely short-ranged, further disproving those possibilities.}. We thus conclude that the presence of electrons at both  $K$ and $K'$ is an artifact of the fact that, for a finite size system, the ground-state is a superposition of states with the two possible senses of valley polarization. This is also consistent with the observation that the maximum value of $n_{\vec k}$ is somewhat less than 1, while for an unpolarized non-interacting system it should equal 2 for all states below the Fermi energy. The occupancy of only one valley naturally explains the observed broken symmetries and the loop currents order, since the persisting current can be estimated as $J_a \approx \pm 2t_1 n \sin(\vec{K}\cdot\hat{e}_a)$, where $\pm$ corresponds to $K$ or $K'$ valley is occupied. Note that the current pattern is translationally invariant, and correspondingly the pattern of flux breaks point group symmetry but not translation symmetry - it is not induced by the PDW order.

The valley polarization can be understood with a simple mean field theory. We propose a trial Hamiltonian
\begin{linenomath}
\begin{align} \label{trial} 
    \hat{H}_{\text{tr}} \equiv \sum_{i, a, \sigma} \left( \tilde{t}_{i, a, \sigma} \hat{c
}_{\vec{r}_i,\sigma}^{\dagger}  \hat{c}_{\vec{r}_i+\hat{e}_a,\sigma} + \text{h.c.}\right) 
\end{align}
\end{linenomath}
to capture the essence of the physics. To simplify the problem, we neglect the constraint on no double occupancy. Further neglecting the weak $\tau$ and $t_2$ terms, we solve the mean field equation  $\tilde{t}_{i,a,\sigma} \approx -t_1 - \frac{J}{2} \langle \hat{c}_{\vec{r}_i+\hat{e}_a,\bar{\sigma}}^{\dagger}  \hat{c}_{\vec{r}_i,\bar{\sigma}}  \rangle- V \langle \hat{c
}_{\vec{r}_i+\hat{e}_a,\sigma}^{\dagger}  \hat{c}_{\vec{r}_i,\sigma}  \rangle$, to the leading order in $n$. We find that the valley-polarized solution, with complex hopping elements $\tilde{t}_{i,a,\sigma} = t'_1 \mathrm{e}^{\mathrm{i}\theta}$ and $\theta \approx  \pm \frac{\sqrt{3}}{4}\frac{J/2+V}{t_1}n$, is always energetically favored. The band structure is renormalized to
\begin{linenomath}
\begin{align}\label{dispersion}
    \epsilon(\Vec{k}) / |t'_1| =  2\cos(k_x+\theta) + 4\cos\left(\frac{\sqrt{3}}{2}k_y\right)\cos\left(\frac{k_x}{2}-\theta\right),
\end{align}
\end{linenomath}
The introduction of the complex phase $\theta$ thus  energetically distinguishes the two valleys by $\frac{9}{2} (J/2+V)n $, the amount of which is sufficiently large to fully valley-polarize the system while keeping the positions of the band minima. This mechanism is similar to that of Stoner magnetism, but here the density of states is divergent at the band bottom so the polarization always occurs in the dilute limit. (Conversely, a finite critical interaction strength is necessary in dimensions $d\geq 2$.)

Therefore, the pairing in each ground state with valley polarization must happen between the two Fermi points located at $k_x =  \Tilde{K}\pm k_F$ and the pair momentum should be twice the center momentum $2\Tilde{K}$. This intra-valley singlet pairing mechanism is distinct from the intra-valley triplet or the inter-valley pairing mechanism proposed for a spin-valley locked system~\cite{Venderleyeaat4698, hsu2017topological}, and is enforced by the broken symmetries. This weak-coupling, mean-field pairing mechanism is complementary to the strong coupling, BEC-type mechanism proposed in Ref.~\cite{PhysRevLett.125.167001}. Furthermore, due to the asymmetry of the non-interacting band structure around its band minima, $\tilde K$ can be calculated to the leading order in $n$:
\begin{linenomath}
\begin{align}
\frac{\Tilde{K}}\pi \approx  \frac{2}{\pi}\arccos\left(\frac{-1}{2\cos
\frac{n\pi}{2}}\right) \approx
\frac K \pi+ 0.45n^2. 
\label{Ktilde}
\end{align}
\end{linenomath}
In Fig.~\ref{fig:Qfit}, we see that the observed $Q$ indeed matches $2\Tilde{K}$ calculated in this way modulo a reciprocal lattice vector, $G=(4\pi,0)$, up to an error of order $\frac{\pi}{N_x}$. This accounts for the incommensurate nature of the PDW. 

{\bf Luther-Emery liquid.} - All the long-distance behaviors we have observed are consistent with a Luther-Emery liquid~\cite{PhysRevLett.33.589}, at least in the two-leg case. For instance, as shown in Fig.~\ref{fig:chargecorrelation}, for electron density $n=0.15$, the oscillatory piece of the charge correlator $C(\Vec{r})\equiv \frac{1}{N_0} \sum_{\Vec{r}_0} \langle n_{\Vec{r}+\Vec{r}_0}n_{\Vec{r}_0}\rangle$ has wave-vector $2k_F=2n\pi$ and exhibits a power-law decay with exponent $\nu_{\text{CDW}}=0.90$, such that the expected relation $\nu_{\text{CDW}}\cdot\nu_{\text{SC}} = 1.08 \approx 1$ is approximately satisfied. The spin correlator $S(\Vec{r})\equiv \frac{1}{N_0} \sum_{\Vec{r}_0} \langle \vec{S}_{\Vec{r}+\Vec{r}_0}\cdot \vec{S}_{\Vec{r}_0}\rangle$ is short-ranged corresponding to a correlation length $\xi\approx 4.36$, as shown in Fig.~\ref{fig:spincorrelation}.
To extract the central charge, we computed the von Neumann entanglement entropy $S_E(x)\equiv -\tr{(\rho_{x}\ln \rho_{x})}$ where $\rho_{x}$ is the reduced density matrix of the subsystem on one side of a cut at $x$. For critical systems in $1+1$ dimensions described by conformal field theory, it has been established~\cite{Calabrese_2004, Fagotti_2011} that for an open boundary system with length $N_x$,
\begin{linenomath}
\begin{align}\label{entropyfit}
S_E(x) =& \,\,\frac{c}{6}\log\left[\frac{4(N_x+1)}{\pi} \sin\left(\frac{\pi(2x+1)}{2(N_x+1)}\right)|\sin q|\right] \nonumber \\
& +\frac{A\sin[q(2x+1)]}{\frac{4(N_x+1)}{\pi}\sin\left(\frac{\pi(2x+1)}{2(N_x+1)}\right)|\sin q|}+B
\end{align}
\end{linenomath}
where $c$ is the central charge, $A$ and $B$ are model-dependent parameters, and $q$ is an adjustable fitting parameter that should approach the Fermi momentum $k_F$ in the thermodynamic limit. We see in Fig.~\ref{fig:centralcharge} that this formula fits well with the observed data. As shown in the inset, the central charge $c \approx 1$ and $q\rightarrow k_F = n\pi$ when $N_x\rightarrow \infty$, which confirms that there is only one gapless mode resulting from filling only one of the two valleys.

{\bf Extensions and speculations:} - With increasing  number of legs, the PDW correlations seem to weaken somewhat, as suggested by the growth of the exponent $\nu_{\text{SC}}$. We speculate that this is an artifact in the few-leg system: at low electron densities, the Fermi pockets are small, so the bands with non-zero $k_y$ and higher energies remain unoccupied for a range of $n$. The effect of adding legs only leads to an increase of the filling of the $k_y=0$ band. In other words, the true scaling process to the two-dimensional limit has not started until the allowed values of $k_y$ become sufficiently closely spaced that additional bands with non-zero $k_y$ cross the Fermi surface. To have more than one band crossing the Fermi surface, the required number of legs scales as $ \sqrt{\frac{4\pi/\sqrt{3}}{n}}$, which is $\sim 12$ for $n=0.05$. Going to larger $n$  could in principle help improve the situation, but this will generally place the system outside of the range of $n$  the PDW occurs - at least for ladders with more than 2 legs. We have not found a range of parameters amenable to DMRG in which the PDW exists and there are multiple Fermi points. Indeed, the results obtained here in the few-leg cases are essentially still one-dimensional; the main insight we obtain from them that could help understanding the 2D case is that valley polarization physics  accounts well for our DMRG results in the 1D case, and this physics should apply as well in two dimensions.

More generally, our findings suggest a new route to PDW order in two or higher dimensions - one that does not rest on complicated strong-coupling physics.  Consider the situation in which, either due to explicit or spontaneous time-reversal symmetry breaking,  a low density of weakly interacting quasiparticles  form a Fermi pocket about a band extremum at a single point in the BZ  (such as the $K$ point in the present context) about which the dispersion  is not symmetric.  Then  from this starting point, it is natural to consider a  pairing instability involving electrons on opposite sides of this pocket.  To the extent that the dispersion relation can be treated as quadratic (i.e. in the effective mass approximation), these states would be perfectly nested, so deviations  from nesting (e.g. trigonal warping) are  small in proportion to a power of the electron density. Consequently, the pairing instability occurs for correspondingly weak couplings, and the resulting state is a PDW with ordering vector or vectors that are likewise continuously varying functions of $n$, as in Eq.~\ref{Ktilde}. This is, in a sense, an orbital version of the original FFLO mechanism.

{\bf Acknowledgement.}
We are grateful to Hong-Chen Jiang, Yi-Fan Jiang, and Patrick Lee for helpful discussions. The DMRG calculations were performed using the ITensor Library~\cite{itensor}. Part of the computational work was performed on the Sherlock cluster at Stanford. SAK was supported, in part, by the National Science Foundation (NSF) under Grant No. DMR2000987.
KSH was supported, in part, by Stanford VPUE through an undergraduate major grant.  HY was supported, in part, by NSFC Grant No. 11825404 at Tsinghua and by the Gordon and Betty Moore Foundations EPiQS through Grant No. GBMF4302 at Stanford.

{\bf Competing interest statement. } Author S.A.K is Editor-in-Chief of npj Quantum Materials. The authors declare no other competing interests.

{\bf Author Contribution.}
KSH and ZH contributed equally to this work. KSH performed the numerical investigation with the help from ZH. All authors analysed the data and contributed to the writing of the paper.

{\bf Data availability. } The authors declare that the data supporting the findings of this study are
available within the paper and its supplementary information files.

{\bf Code availability. } The codes implementing the calculations of this study are available from the corresponding author upon request.

\bibliographystyle{apsrev4-1}

\onecolumngrid

\onecolumngrid
\setcounter{equation}{0}
\setcounter{figure}{0}
\renewcommand{\theequation}{S\arabic{equation}}
\renewcommand{\figurename}{Supplementary Figure}

\section*{Supplementary Discussion}

\subsection{The key observables with another set of parameters}
In Supplementary Figure \ref{fig:A} we show some key data for another set of parameters. We confirm that, although the parameters have been changed significantly, we still observe  strong PDW correlations with exponent $\nu_\text{SC} = 1.83$ and thus a divergent susceptibility, as shown in Supplementary Figure  \ref{fig:A_pdw}. The valley polarization picture still applies, as indicated by the width of the Fermi pockets $2k_\text{F}=2n\pi$ shown in Supplementary Figure  \ref{fig:A_nk}. 
\begin{figure}[h]
	\centering
	\subfigure[]{\label{fig:A_pdw}\includegraphics[width=0.4\linewidth]{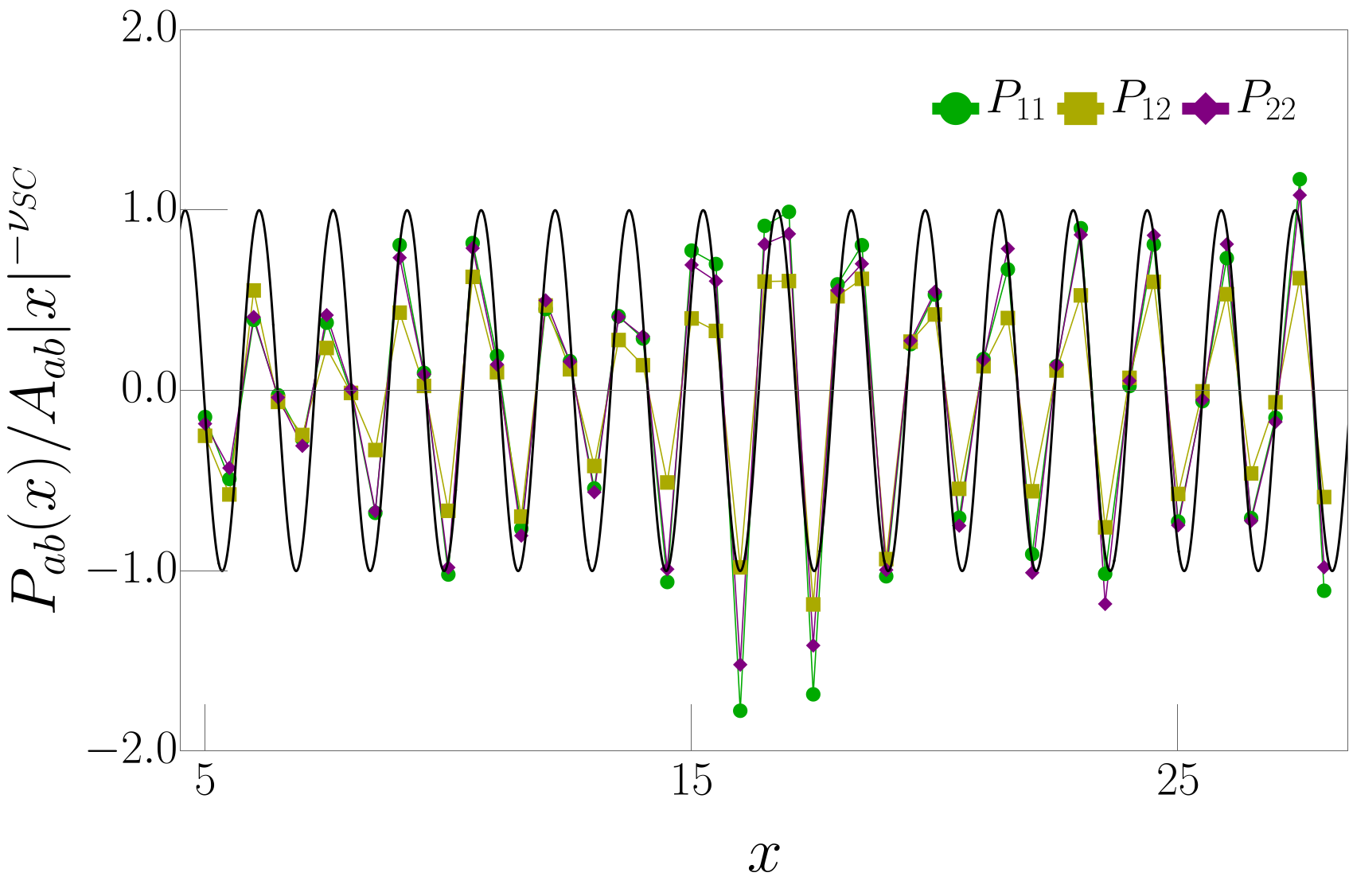}}
	\subfigure[]{\label{fig:A_nk}\includegraphics[width=0.24\linewidth]{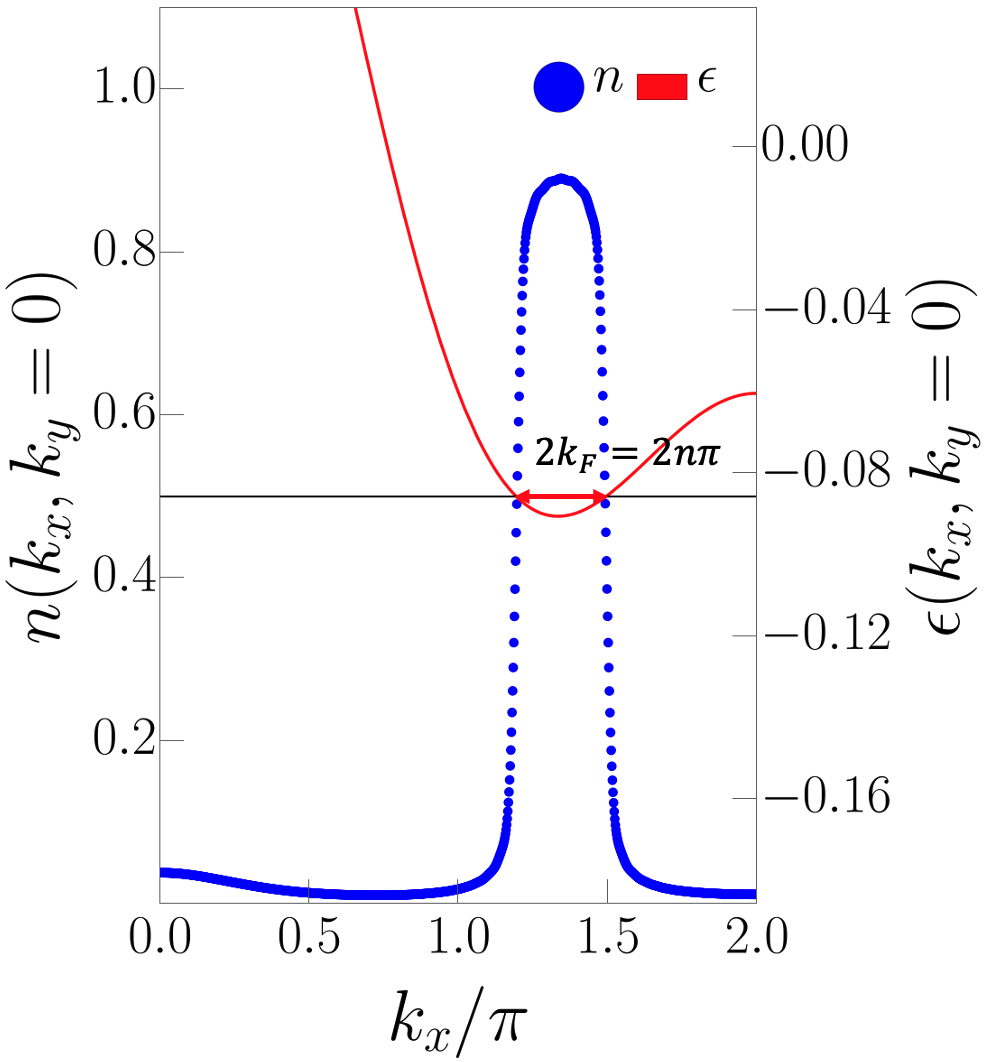}}
	\caption{\label{fig:A} Data for $t=-1$, $\omega_D=2.857$, $U_\text{e-e}=28$, and $U_\text{e-ph}=20$, which leads to $J \approx 0.155$, $ V/J \approx 0.780$, $t_1/J \approx -0.195$, $t_2/J \approx 0.0250$ and $\tau/J \approx 0.0238$ in the effective model. The truncation error is $4 \times 10^{-7}$. (a) The pair-pair correlator normalized by a power-law decay function $A_{ab}x^{-\nu_\text{SC}}$ with $\nu_{SC}=1.83$ for a two-leg cylinder of length $N_x=100$ and electron density $n=0.15$. The black reference curve is $\cos(Qx+\phi)$ with $Q=1.315\pi$ extracted from the structure factor. (b) The occupation number in momentum space of a two-leg cylinder with length $N_x=100$ and electron density $n=0.15$. The Fermi points match those of the non-interacting band with width $2k_\text{F}=2n\pi=0.3\pi$. The interval $k_x \in (-2\pi,0)$ mirrors the range shown.}
\end{figure}

\subsection{The key observables without $t_2$ and $\tau$ terms}
In Supplementary Figure \ref{fig:B}, we show some key data for the same parameters as in main text but setting $t_2=\tau=0$, i.e. the pure $t$-$J$-$V$ model. We confirm that, without the $t_2$ and $\tau$ terms, the PDW correlation weakens in comparison to the original effective model,  but the exponent $\nu_\text{SC} = 1.90<2$ still yields a divergent susceptibility, as shown in Supplementary Figure \ref{fig:B_pdw}. The occupation number and thus the valley polarization picture is basically unaffected, except that there is no longer a small fraction of occupation weight being shifted to the vicinity of zero momentum, as shown in Supplementary Figure \ref{fig:B_nk}. 
\begin{figure}[h]
	\centering
	\subfigure[]{\label{fig:B_pdw}\includegraphics[width=0.4\linewidth]{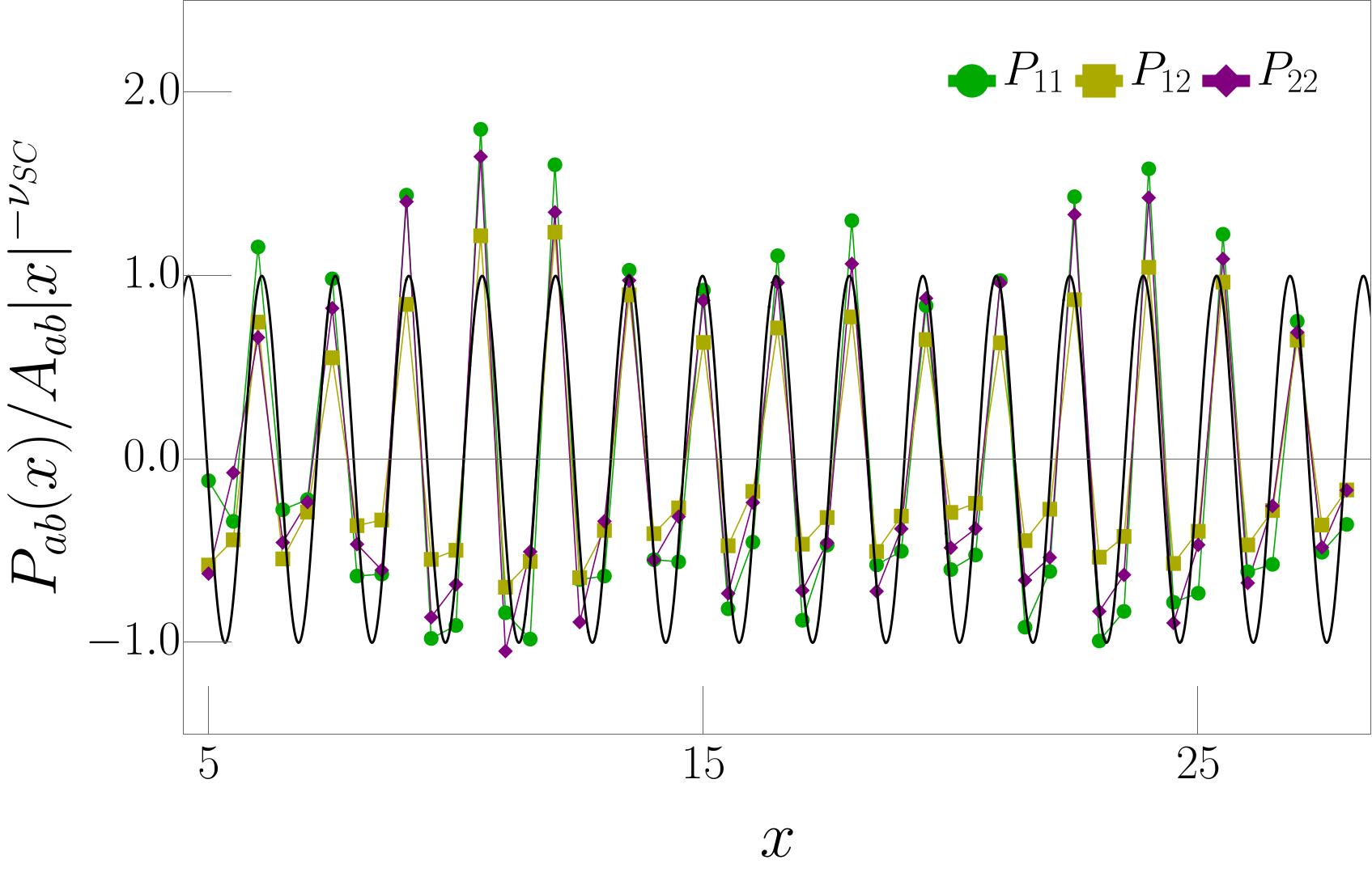}}
	\subfigure[]{\label{fig:B_nk}\includegraphics[width=0.24\linewidth]{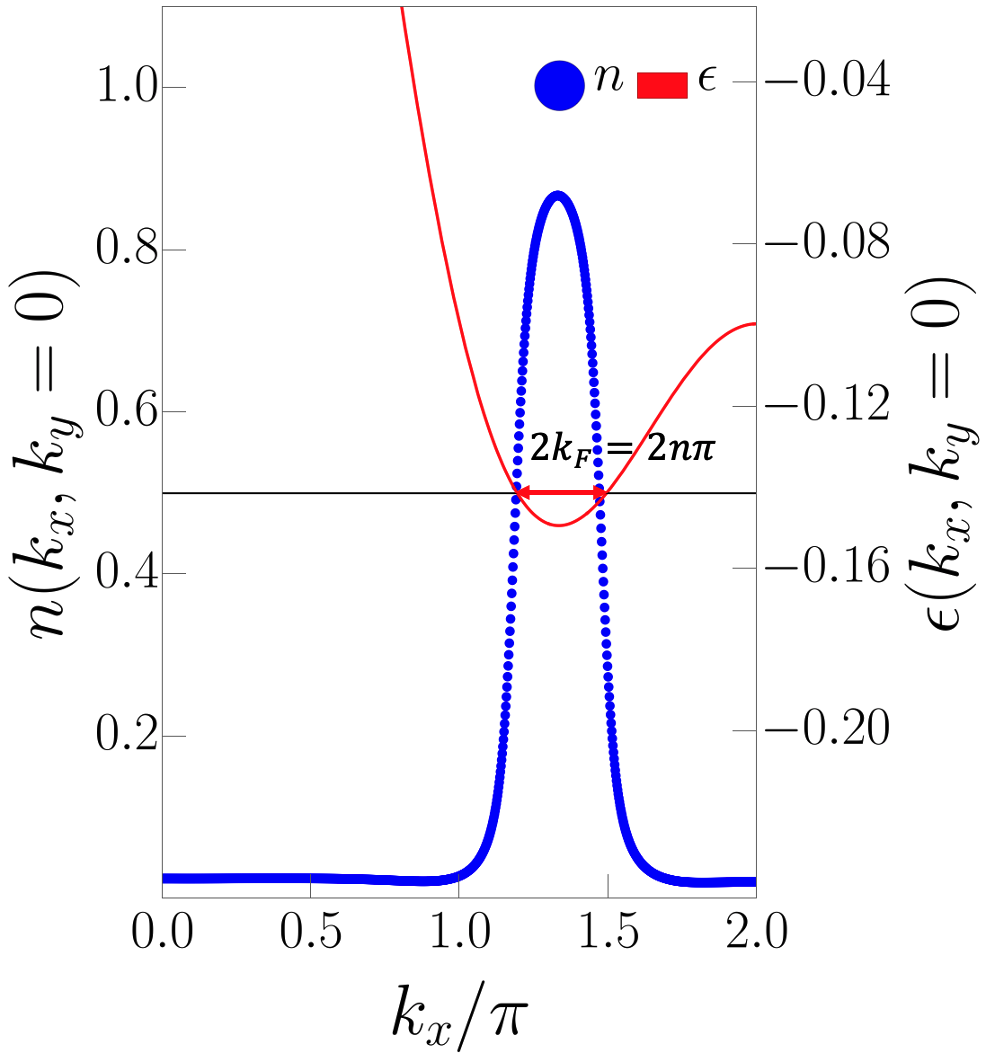}}
	\caption{\label{fig:B} Data for the same parameters as in main text but setting $t_2=\tau=0$ ($J \approx 0.208$, $ V/J \approx 0.666$, $t_1/J \approx -0.240$). The truncation error is  $4 \times 10^{-7}$. (a) The pair-pair correlator normalized by a power-law decay function $A_{ab}x^{-\nu_\text{SC}}$ with $\nu_{SC}=1.90$ for a two-leg cylinder of length $N_x=240$ and electron density $n=0.15$. The black reference curve is $\cos(Qx+\phi)$ with $Q=1.348\pi$ extracted from the structure factor. (b) The occupation number in momentum space of a two-leg cylinder with length $N_x=240$ and electron density $n=0.15$. The Fermi points match those of the non-interacting band with width $2k_\text{F}=2n\pi=0.3\pi$. The interval $k_x \in (-2\pi,0)$ mirrors the range shown.}
\end{figure}

\subsection{The current operators and the scaling of current-current correlation}
The current operator can be derived from the time derivative of the density operator:
\begin{align}
   \hat{\Dot{n}}_i &= \mathrm{i} [\hat{H},\hat{n}_i] \nonumber\\
    &= \mathrm{i} t_1 \sum_{\langle i,j\rangle,\sigma}\left( \hat{c}_{i,\sigma}^{\dagger}  \hat{c}_{j,\sigma} - \hat{c}_{j,\sigma}^{\dagger}  \hat{c}_{i,\sigma}\right) + \mathrm{i}t_2 \sum_{\langle i, m, j\rangle,\sigma}(1-2\hat n_m) \left( \hat{c}_{i,\sigma}^{\dagger}  \hat{c}_{j,\sigma} - \hat{c}_{j,\sigma}^{\dagger}  \hat{c}_{i,\sigma}\right)  +\mathrm{i}(\tau+2t_2)\sum_{\langle i, m, j\rangle} \left( \hat{s}^{\dagger}_{im} \hat{s}_{mj}  -  \hat{s}^{\dagger}_{jm} \hat{s}_{mi}  \right)
\end{align}
where the last two terms will contribute to the current on both bond $\langle im\rangle$ and $\langle mj\rangle $. Therefore, the current operator on bond $\langle ab\rangle $ can be written as:
\begin{align}
   \hat{J}_{b\rightarrow a} 
    &= \mathrm{i} t_1 \sum_{\sigma}\left( \hat{c}_{a,\sigma}^{\dagger}  \hat{c}_{b,\sigma} - \hat{c}_{b,\sigma}^{\dagger}  \hat{c}_{a,\sigma}\right) \nonumber \\
& \ \ \  + \mathrm{i}t_2 \sum_{\langle a, b, c\rangle,\sigma} (1-2\hat n_b)\left( \hat{c}_{a,\sigma}^{\dagger} \hat{c}_{c,\sigma} - \hat{c}_{c,\sigma}^{\dagger}  \hat{c}_{a,\sigma}\right)  + \mathrm{i}t_2 \sum_{\langle  c, a, b\rangle,\sigma} (1-2\hat n_a)\left( \hat{c}_{c,\sigma}^{\dagger} \hat{c}_{b,\sigma} - \hat{c}_{b,\sigma}^{\dagger}  \hat{c}_{c,\sigma}\right)  \nonumber \\
& \ \ \ +\mathrm{i}(\tau+2t_2)\sum_{\langle a, b, c\rangle} \left( \hat{s}^{\dagger}_{ab} \hat{s}_{bc}  -  \hat{s}^{\dagger}_{cb} \hat{s}_{ba} \right)  +\mathrm{i}(\tau+2t_2)\sum_{\langle c,a, b\rangle} \left( \hat{s}^{\dagger}_{ca} \hat{s}_{ab}  -  \hat{s}^{\dagger}_{ba} \hat{s}_{ac} \right) 
\end{align}

The $z$ component of the spin current operator on bond $\braket{ab}$ can be written as $\hat{J}^s_{b \rightarrow a}=\hat{J}^{\uparrow}_{b \rightarrow a}-\hat{J}^{\downarrow}_{b \rightarrow a}$ with:
\begin{align}
   \hat{J}^{\sigma}_{b\rightarrow a} 
    &= \mathrm{i} t_1 \left( \hat{c}_{a,\sigma}^{\dagger}  \hat{c}_{b,\sigma} - \hat{c}_{b,\sigma}^{\dagger}  \hat{c}_{a,\sigma}\right) \nonumber \\
& \ \ \  + \mathrm{i}t_2 \sum_{\langle a, b, c\rangle} (1-2\hat n_b)\left( \hat{c}_{a,\sigma}^{\dagger} \hat{c}_{c,\sigma} - \hat{c}_{c,\sigma}^{\dagger}  \hat{c}_{a,\sigma}\right)  + \mathrm{i}t_2 \sum_{\langle  c, a, b\rangle} (1-2\hat n_a)\left( \hat{c}_{c,\sigma}^{\dagger} \hat{c}_{b,\sigma} - \hat{c}_{b,\sigma}^{\dagger}  \hat{c}_{c,\sigma}\right) 
\end{align}
The singlet hopping term does not contribute to the spin current.

In Supplementary Figure \ref{fig:SJpeaks} we plot the magnitude of the zero-momentum current-current structure factor at various lengths of the two-leg cylinder. We find that its magnitude scales linearly with the system size, signifying  long-range order and thus the existence of persisting currents in the ground states. In Supplementary Figure \ref{fig:spincurr}, we confirm the exponential decay of the spin  current correlations at long distances, which rules out the possibility of spin-valley locked polarization.

\begin{figure}[h]
    \centering
	\subfigure[]{\label{fig:SJpeaks}\includegraphics[width=0.45\linewidth]{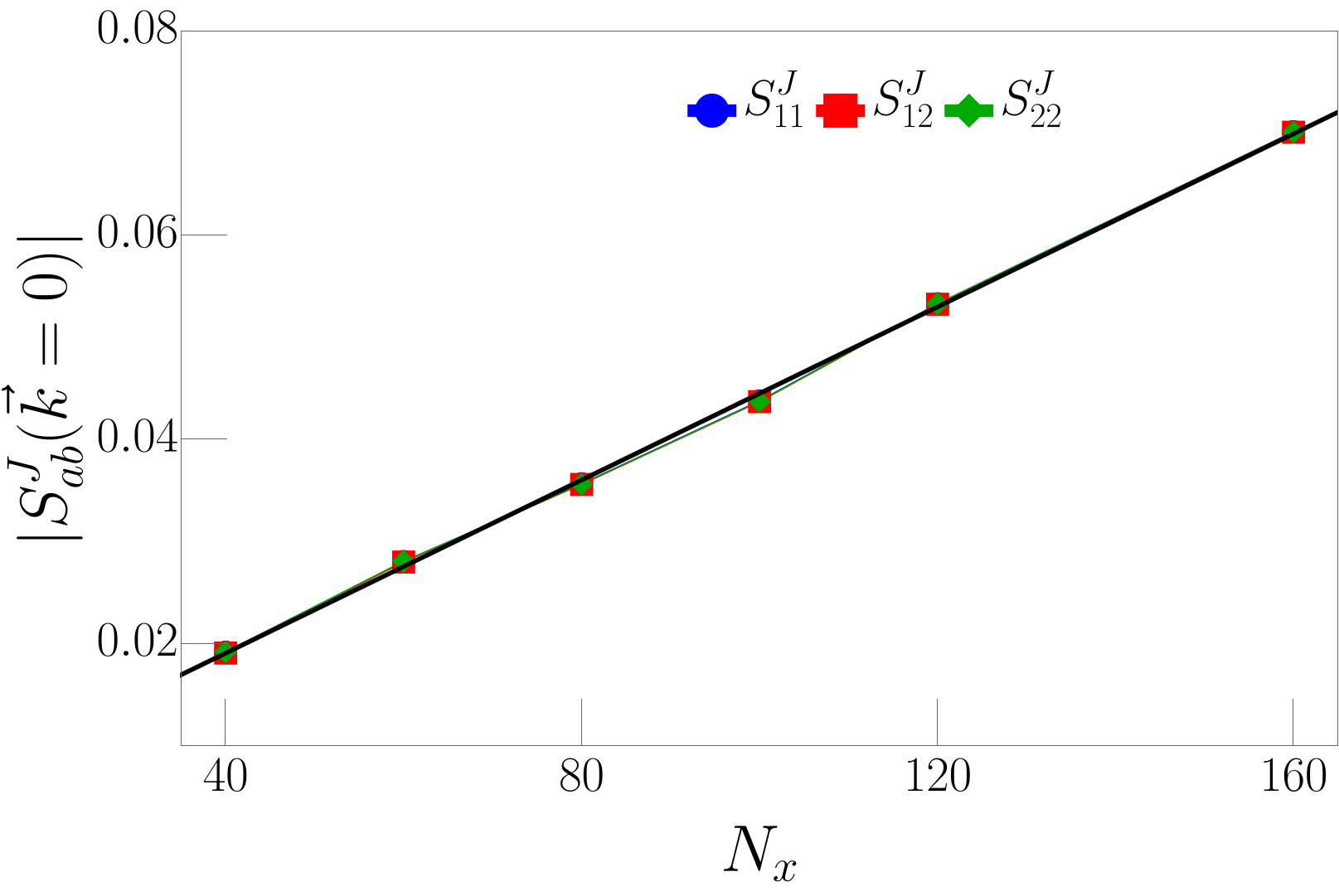}}
	\subfigure[]{\label{fig:spincurr}\includegraphics[width=0.45\linewidth]{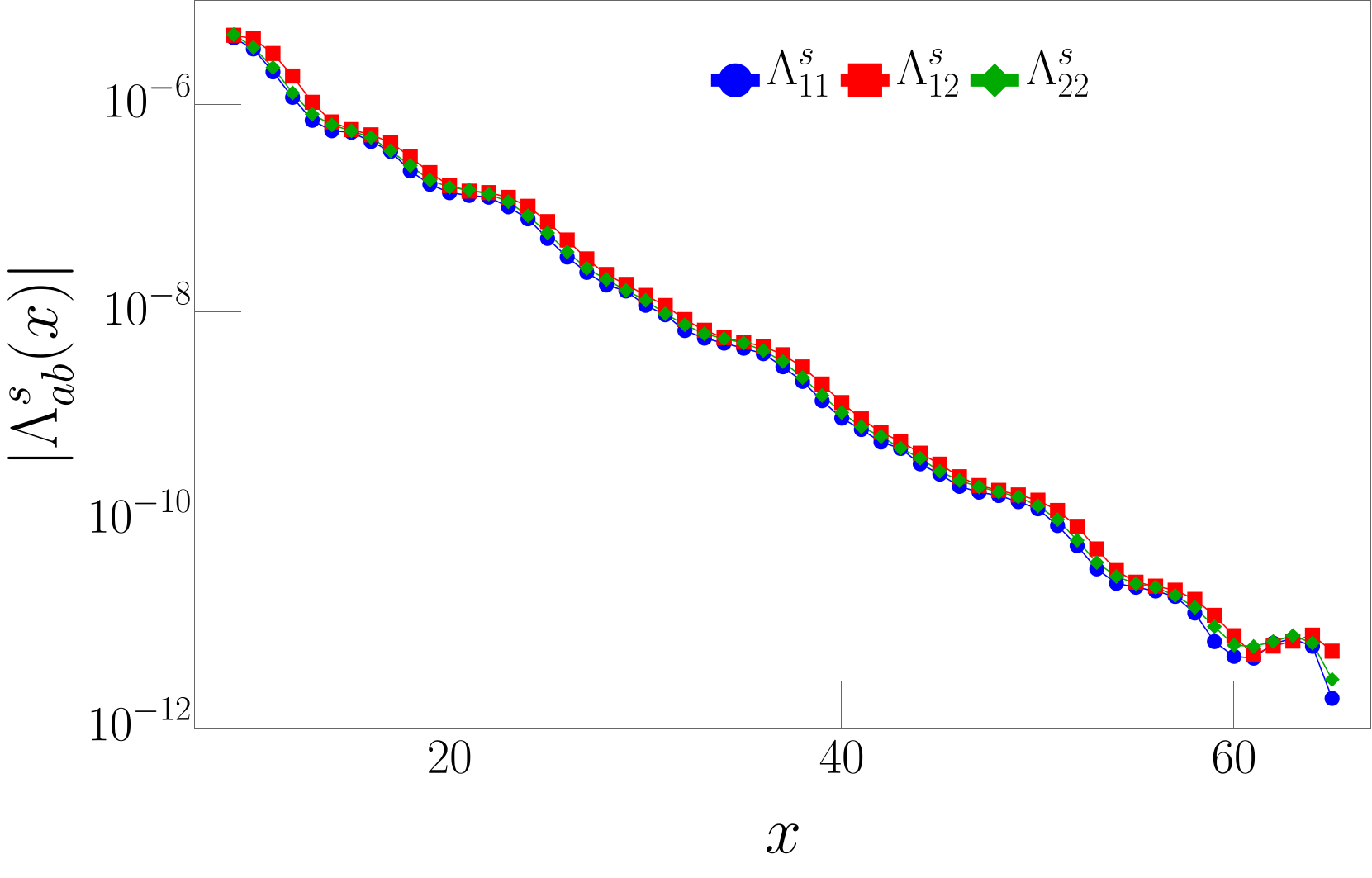}}
	\caption{\label{fig:C} For the same parameters as in main text, with electron density $n=0.15$ and truncation error less than $5 \times 10^{-7}$, (a) Magnitude of current-current structure factor $|S^J_{ab}(\mathbf{k}=0)|\equiv |\frac{1}{N}\sum_{\mathbf{r}_i,\mathbf{r}_j} \braket{\mathbf{J}_a(\mathbf{r}_i) \cdot \mathbf{J}_b(\mathbf{r}_j)}|$ at various lengths of the two-leg cylinder with electron density $n=0.15$. The black curve is a linear fit $AN_x+B$ which yields $A/t_1^2=0.17$. (b) The spin current correlator for a two-leg cylinder of length $N_x=240$ and electron density $n=0.15$ where $\Lambda^s_{ab}(\mathbf{r}) \equiv \frac{1}{N_0} \sum_{\mathbf{r}_0} \langle \mathbf{J}^s_a(\mathbf{r}+\mathbf{r}_0)\cdot\mathbf{J}^s_b(\mathbf{r}_0)\rangle$.}
\end{figure}

\subsection{Fermi density for four-leg case}
To verify that the evidence of valley polarization still shows up for more-leg case, we plot the Fermi density for four-leg case in Supplementary Figure \ref{fig:4leg}. We see that the Fermi surface size is still twice what would be for non-interacting system, and the Fermi density inside the pocket is slightly less than one.
\begin{figure}[h]
	\centering
	\subfigure[]{\label{fig:4leg_nk}\includegraphics[width=0.2\linewidth]{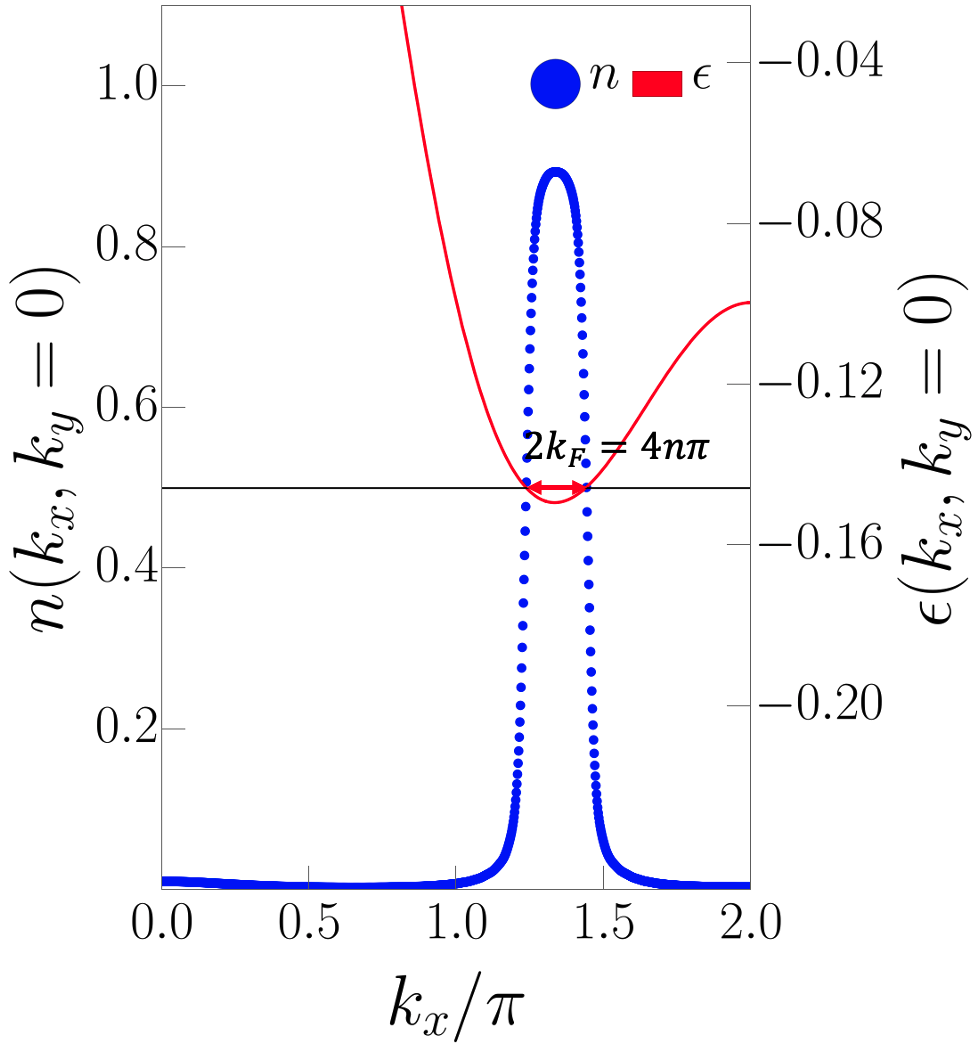}}
	\caption{\label{fig:4leg} The occupation number in momentum space of a four-leg cylinder with length $N_x=100$ and electron density $n=0.05$. The Fermi points match those of the non-interacting band with width $2k_\text{F}=4n\pi=0.2\pi$. The interval $k_x \in (-2\pi,0)$ mirrors the range shown.}
\end{figure}

\subsection{Truncation error extrapolation}
In Supplementary Figure \ref{fig:extrap}, we show the truncation error extrapolation data for some main results shown in the main text for the two-leg case. Each data point is extrapolated to zero truncation error by a quadratic fit (marked as $\epsilon=0$), utilizing data collected with six truncation errors ranging from $3\times 10^{-6}$ to $1\times 10^{-7}$. We have also checked that the extrapolated results do not differ significantly from the result obtained with a very small truncation error. For more leg cases, we did the same extrapolation for all results shown in the main text. For those ``sanity check'' simulations with very small truncation error, the maximum bond dimension is $3125$ and the simulation took $126$ cpu hours for 4-leg system $N_x=100$, $n=0.05$ and truncation error $7 \times 10^{-11}$; the maximum bond dimension is $2602$ and the simulation took $260$ cpu hours for 6-leg system $N_x=60$, $n=0.05$ and truncation error $7 \times 10^{-10}$. We remark that, although the bond dimensions look relatively small (due to the smallness of Hilbert space in dilute limit), all the simulations are still quite time-consuming, due to the complicated form of the Hamiltonian.

\begin{figure}[h]
	\centering
	\subfigure[]{\label{fig:pdwextrap}\includegraphics[width=0.41\linewidth]{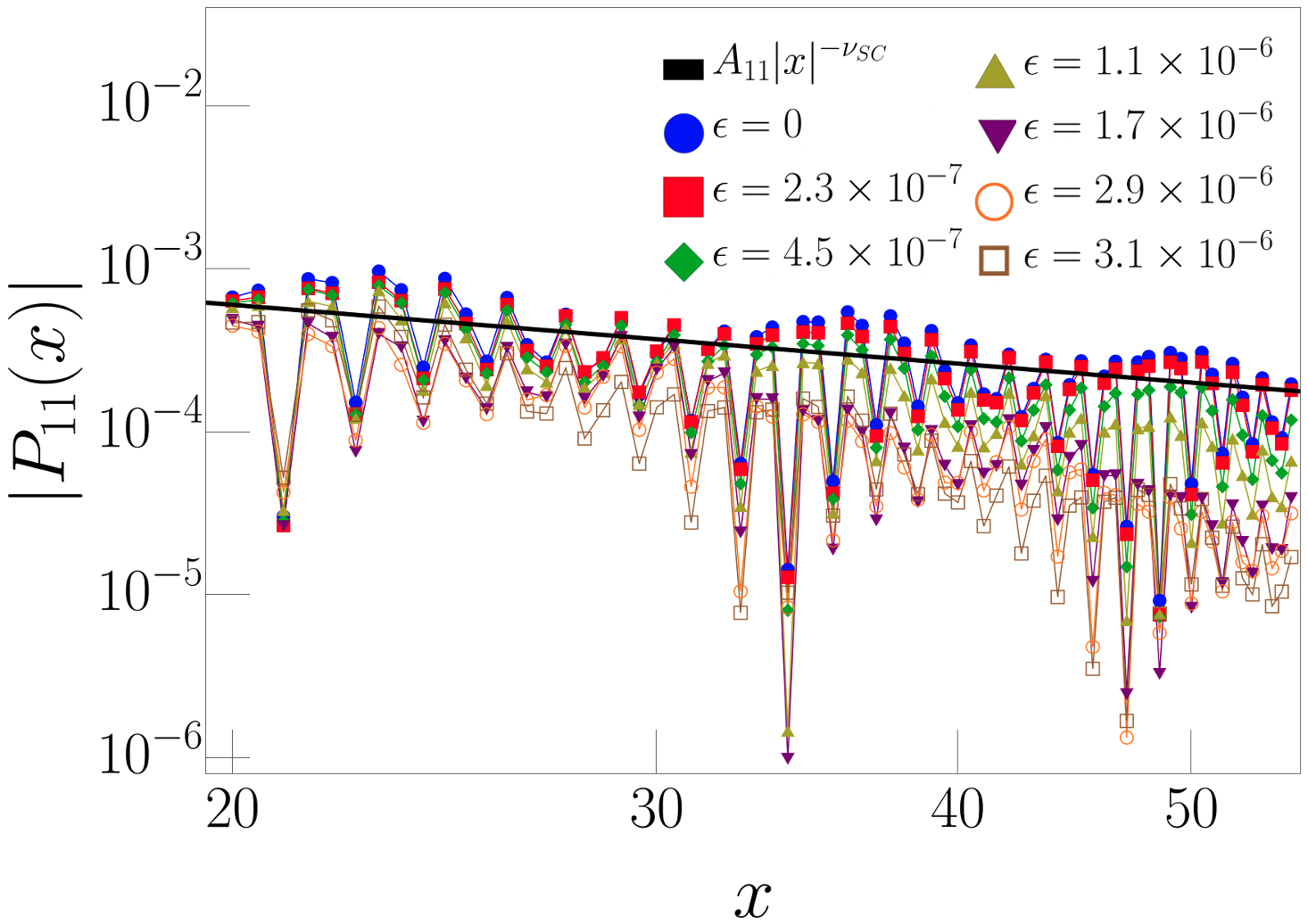}}
	\subfigure[]{\label{fig:cdwextrap}\includegraphics[width=0.30\linewidth]{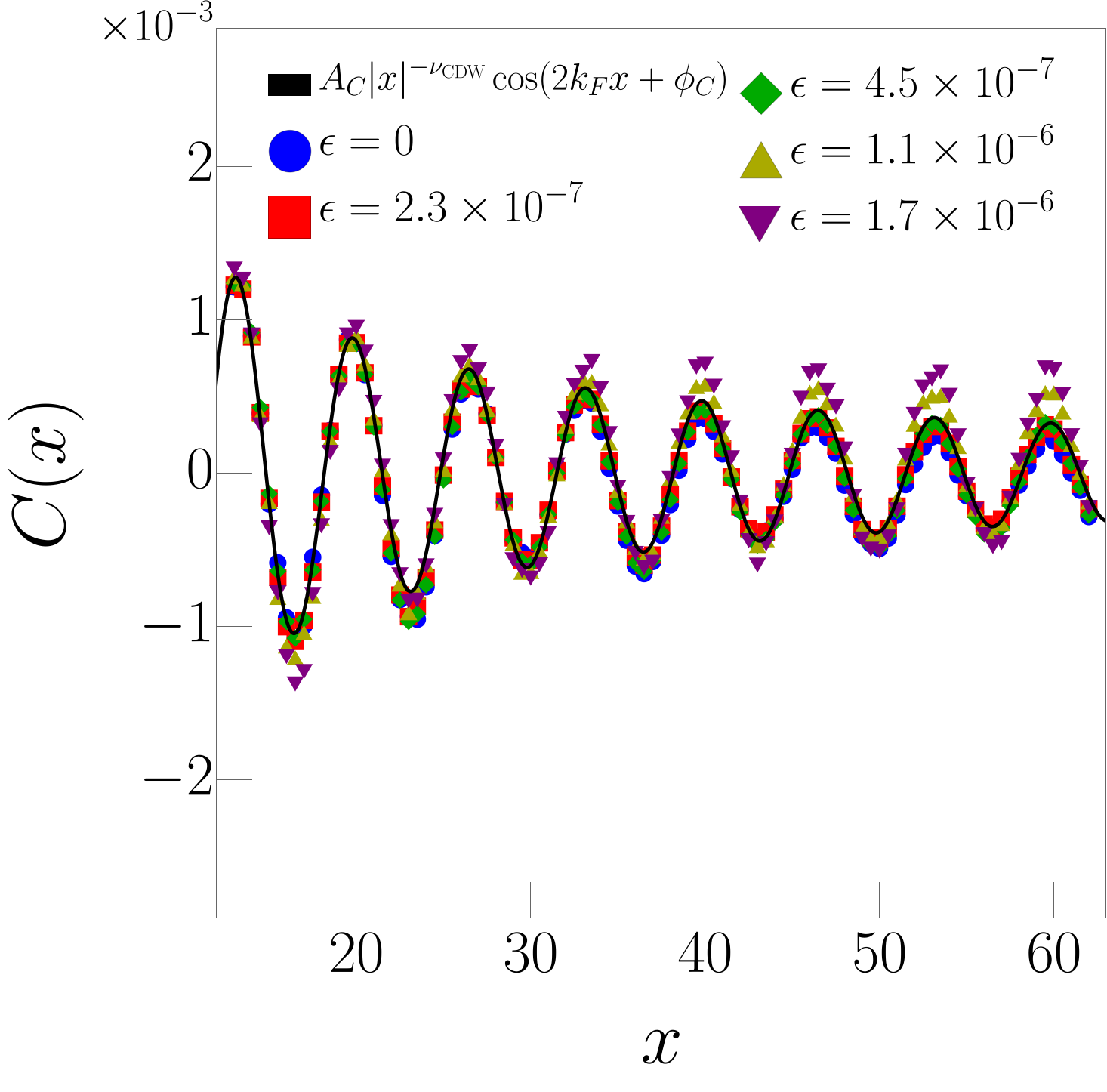}}
	\subfigure[]{\label{fig:sdwextrap}\includegraphics[width=0.22\linewidth]{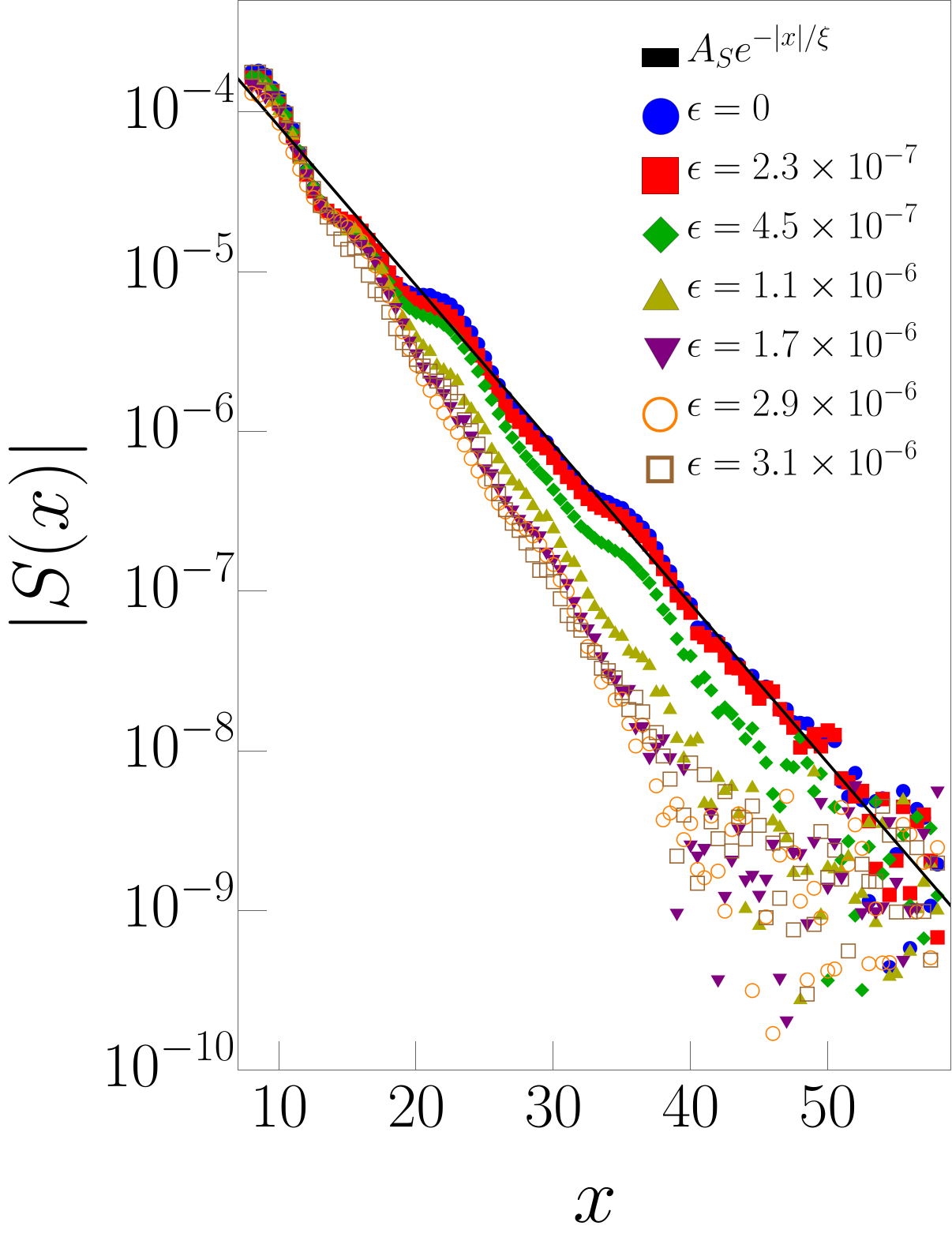}}
	\caption{\label{fig:extrap} For the same parameters as in main text for two-leg ladder with electron density $n=0.05$,  truncation error extrapolation for (a) pair-pair correlator $P_{11}$, (b) CDW correlation, (c) SDW correlation. ``$\epsilon = 0$'' labels the data point that is extrapolated to zero truncation error by a quadratic fit.}
\end{figure}

\end{document}